\documentclass[11pt]{article}

\title{Rare Events Analysis and Computation for  Stochastic Evolution  of Bacterial Populations}
\author{Yingxue Su\thanks{University of Houston, Department of Mathematics, yingxue@math.uh.edu} \and 
Brett Geiger\thanks{High Point University, Mathematics, bgeiger@highpoint.edu}\and 
Ilya Timofeyev\thanks{University of Houston, Department of Mathematics, ilya@math.uh.edu} \and 
Andreas Mang\thanks{University of Houston, Department of Mathematics, andreas@math.uh.edu} \and 
Robert Azencott\thanks{University of Houston, Department of Mathematics, razencot@math.uh.edu}}

\date{\today}

\usepackage{graphicx}
\usepackage{booktabs}
\usepackage{amsmath}
\usepackage{caption}
\usepackage{subcaption}
\usepackage{amsthm}
\usepackage{amsfonts}
\usepackage{fullpage}
\usepackage[colorlinks=true, linkcolor=blue, citecolor=blue]{hyperref}
\usepackage{bm}
\usepackage{scalerel}

\usepackage{amsthm}

\newtheorem{definition}{Definition}[section]
\newtheorem{theorem}{Theorem}[section]
\newcommand{\opt}{\star}
\newcommand{\eps}{\varepsilon}
\newcommand{\tabref}[1]{Table~\ref{#1}}

\newcommand{\secref}[1]{Section~\ref{#1}}
\newcommand{\apxref}[1]{Appendix~\ref{#1}}

\newcommand{\figref}[1]{Fig.~\ref{#1}}

\newcommand\MTR{M}
\newcommand\Rt{R}
\newcommand\bR{\mathbb{R}}
\newcommand\bE{\mathbb{E}}
\newcommand\bN{\mathbb{N}}

\newcommand\btheta{\bm{\theta}}
\newcommand\bh{\mathbf{H}}
\newcommand\bhh{\mathbf{h}}
\newcommand\bH{\mathbf{H}}

\newcommand\calM{\mathcal{M}}
\newcommand\calS{\mathcal{S}}
\newcommand\calH{\mathcal{H}}

\newcommand\MLP{L}
\newcommand\mut{u}
\newcommand\hatH{\widehat{H}}
\newcommand{\argmin}{\operatornamewithlimits{argmin}}

\begin{document}

\maketitle

\begin{abstract}
In this paper, we develop a computational approach for computing most likely trajectories describing rare events that correspond to the emergence of non-dominant genotypes. This work is based on the large deviations approach for discrete Markov chains describing the genetic evolution of large bacterial populations. We demonstrate that a gradient descent algorithm developed in this paper results in the fast and accurate computation of most-likely trajectories for a large number of bacterial genotypes. We supplement our analysis with extensive numerical simulations demonstrating the computational advantage of the designed gradient descent algorithm over other, more simplified, approaches.
\end{abstract}


\section{Introduction}

Many biological systems can be modeled as random complex systems with a large number of individual interacting agents. The genetic evolution of large bacterial or viral populations is one such example. Moreover, in many applications, random mutations play an important role in describing rare events,  such as the emergence of non-dominant genotypes with new biological properties.  Examples of such rare events have been observed in long genetic experiments of \textit{Escherichia coli}. For instance,  variation of resistance levels in bacterial populations without exposure to antibiotics~\cite{lmls2019}, the emergence of genotypes with lower fitness, but with a higher  degree of adaptation or evolvability~\cite{Barrick2010,Woods2011}, or the emergence of wider cell sizes but later reverting to the original cell shape~\cite{nkrumah-lenski2021} are only a few such examples.

Large-deviations theory provides a unified and efficient framework for studying rare events in a wide range of random models, such as Markov chains ({\bf MC}s), Gaussian processes, stochastic differential equations, etc (we refer for instance to~\cite{dembo,varadhan,ventcell,azen77,azen80,azenLDT2012,bgte2016, Grafke2017,greve2019} for examples). For instance, for random evolution of large populations, interesting large deviations results were obtained in~\cite{champ6,champ01,champ8,champ02}. These works study Darwinian evolution of asexual populations in the vector space of phenotypic traits that are transmitted to offspring with rare variations due to gene mutations. In this model, competition for limited resources forces a roughly periodic selection.  In the present work, we focus on rare random genetic evolutionary events in  bacterial populations, such as the emergence of a low-fitness genotype that reaches high frequency in the population.  Computing the most likely evolutionary paths realizing such rare genetic events is a complicated mathematical problem, even for simplified stochastic models of bacterial population evolution. The stochastic genetic evolution of large bacterial populations observed in long-term laboratory experiments (see,   e.g., \cite{azencoop,heger,Gordo2013, Illinworth2012, Simon2013,Ross1996}) has typically been modeled by simplified MC models (see \secref{s:s_mod}), which are fitted to experimental data.  To analyze rare events for these types of MC models, we developed~\cite{azencott2018rare, geiger_thesis} a large deviations theoretical framework summarized in \secref{sec3}. 

The numerical computation of the most likely evolutionary paths realizing rare genetic events involves an optimization problem, and thus, is computationally challenging for straightforward search algorithms as soon as the number of genotypes exceeds four. In the present paper, we develop and test several algorithms for the numerical computation of the most likely genetic evolution trajectory connecting the initial population histogram $H$ and a terminal histogram $G$ (see also \cite{su_thesis}).  We demonstrate that the gradient decent algorithm developed in this paper (see \secref{sec5}) is capable of computing the most likely evolutionary trajectory for a large number (up to twenty) of interacting genotypes. In particular, we present a computational example for ten genotypes and compare computing times with more straightforward approaches.

\paragraph*{Laboratory Experiments on Bacterial Genetic Evolution.} For bacteria, such as \emph{Escherichia coli}, genetic evolution has been explored in many long-term laboratory experiments (see, e.g., \cite{Lenski1991,cooper1,heger,len94a,Fox2015,Barrick2013,Barrick2010,Cooper2001, Deatherage2017,Woods2011,Levy2015,Plank1979,rice}. For instance, in~\cite{heger,azencoop}, on each day $t$ a bacterial population with initial large size $N$ cells grows freely during the day until the daily dose of nutrients is exhausted. After nutrient exhaustion, the cell population remains dormant until the end of the day. Therefore, daily growth duration is practically determined by the (fixed) daily dose of nutrients and hence can be considered fixed throughout the experiment. Growth of each sub-population with a particular fixed genotype is determined by the corresponding growth rate, which is related to the fitness of this genotype. In addition, rare random mutations can also occur. At the end of the day $t\in\mathbb{N}$, the population size becomes a large multiple of $N$, and one selects by dilution a random sub-sample of roughly $N$ cells, which becomes the initial population on the day $t+1$. Thus, the bacterial population at the beginning of each day is roughly the same size, $N$. Typical values for $N$ can range from $10^5$ to $10^8$. For $g\in\mathbb{N}$ genotypes, one records (daily if feasible) frequencies of cells with each genotype and thus, each population is characterized by its population histogram $H_t=[H_t(1), H_t(2), \ldots, H_t(g)]\in \mathbb{R}^g$ of genotype frequencies $H_t(j)$, $j=1,\ldots,g$. For the colony of $j$-cells (i.e., cells with genotype $j$), daily growth by cell divisions roughly multiplies its initial size by a fixed growth factor $F_j \gg 1$. We assume that genotypes are ordered by increasing fitness so that $F_1 < F_2< \ldots < F_g$. Thus, the genotype $g$ is called \emph{dominant}, and its frequency $H_t(g)$ tends to  1 for large $t$. At each $j$-cell division, the genotype $j$ is typically inherited by the two daughter cells, unless a very rare random genotype mutation from $j$ to $k \neq j$ occurs. Mutations approximately follow a Poisson distribution with a very small mutation rate, typically ranging from $10^{-9}$ to $10^{-6}$.

\paragraph*{Fixation of Non-Dominant Genotypes and Rare Genetic Events.} For any \emph{non-dominant} genotype $j<g$ initially absent in the population, a $j$-cell colony may emerge for the first time on day $t$, due to random mutations. However, for large populations of size $N$, this new $j$-cell colony has an extremely small probability of reaching \emph{fixation} at some later day $t+\tau$, i.e., of reaching a high frequency $H_{t+\tau}(j)$ (e.g., higher than $40\%$). Indeed, this probability vanishes at an exponential rate as $N$ increases. To enable a precise analysis of rare genetic events such as fixation of non-dominant genotypes we have developed (see \cite{azencott2018rare}) a \emph{large deviation theory} framework underlying the stochastic genetic evolution of large bacterial populations. In our stochastic models, the initial population at the beginning of each day $t$ has the same size $N$ due to random sub-sampling selection at the end of the previous day. Consider any fixed time $T >0$.  The random genetic evolution of the bacterial population over $T$ days is then a random sequence of histograms $\mathbf{H}= \{H_1, \ldots, H_T\}$, characterized by a MC over the set $\calH \subset \bR^g$ of all histograms of dimension $g$. This MC path space $\Omega_T$ is the set of all possible sequences $\mathbf{H}= \{H_1, \ldots, H_T\}$ of population histograms. For large $N > 10^5$ and fixed initial histogram $H$, the probability distribution $P_N$ of the random paths $\mathbf{H}$ starting at $H_1 = H$ is highly concentrated around a single path, namely the unique \emph{mean  trajectory} $\mathbf{\MTR} = \{\MTR_t, \, t=1,2,\ldots,T\}$ starting at $H$ (i.e., $\MTR_1= H$). 

In a large deviation theory framework, a key step is to determine the \emph{rate functional} \[\lambda:\Omega_T \to \bR^+\] which controls the probabilities of rare genetic events (we define $\lambda$ in \eqref{e:firstlambda}). For each population trajectory $\bH \in \Omega_T$ starting at $H_1=H$,  we have computed the limit
\[
\lim_{N \to \infty} \frac{1}{N} \log P_N(\bh) = - \lambda(\bh) \leq 0.
\]

\noindent Whenever $\bh$ is not identical to the mean trajectory $\mathbf{\MTR}$ starting at $H_1=H$, one has $\lambda(\bh) >0$ and the probability $P_N(\bh)$ is a vanishing exponential of the order of $\exp({-N\lambda(\bh)})$.

\paragraph*{Contributions.} Given any two population histograms $H$ and $G$,  the \emph{most likely} random population path $\mathbf{H} \in \Omega_T$ such that $H_1=H$ and $H_T=G$ is the deterministic path $\bh^\opt$  minimizing the cost function $\lambda(\bh) $ over all paths  $\bh \in \Omega_T$ such that $H_1=H$ and $H_T= G$. The \textit{\textbf{main focus of this paper is developing and testing practical algorithms enabling computation of the most likely path}} $\bh^\opt$ connecting two given histograms $H$ and $G$ in $T$ steps. This task can be quite challenging for straightforward algorithms when the number of genotypes $g$ exceeds four. This problem has several natural analogies with the computation of a geodesic connecting two points on a smooth Riemannian manifold. In particular, we have identified a recursive reverse-time equation for computing the optimal trajectory $\bh^\opt$ given the final and penultimate histograms. Since the penultimate histogram is not known a priori, this leads to an optimization problem with respect to the penultimate histogram.
We present several algorithms to numerically compute $\bh^\opt$ for $3\leq g \leq 20$. 
We also would like to point out that computing these rare paths is crucial for estimating fixation probabilities by \emph{importance sampling} simulations. For instance, after the most likely path has been computed, one can develop Monte-Carlo importance sampling simulations for trajectories which are close to the most-likely path
and, thus, efficiently estimating probabilities of rare events (e.g. probability of reaching the final histogram).
This approach was implemented in \cite{su_thesis}.

The code is available at \cite{coderepo}.

\paragraph*{Manuscript Organization.}
In the next section, we describe the MC model and discuss the three daily stages for bacterial populations: growth, mutations, and dilution. In Section \ref{sec3} we discuss the main results from the large deviations theory. In particular, we introduce the one-step cost function \eqref{e:costformula} and the formula for the reverse computation of the most likely trajectory \eqref{e:reverse}.
We present the two simpler algorithms for the computation of the most likely evolutionary trajectories and discuss their efficiency
in Section \ref{s:brute_algo}.
The gradient descent algorithm is presented in Section \ref{sec5}.
Conclusions are presented in Section \ref{sec6} and Appendix
\ref{sec:ap1}, \ref{ap:params}, and \ref{s:gradient}
summarizes the main notation used in this paper, computational parameters, and the gradient descent formulas, respectively.

\section{Simplified Genetic Evolution Model for Bacterial Populations}
\label{s:s_mod}

The MC model of population evolution studied here, as well as in~\cite{azencott2018rare}, has been applied to emulate the stochastic evolution of large  bacterial populations observed in long-term laboratory experiments~\cite{azencoop}. We assume that there is a fixed number of possible genotypes denoted as $\{1, 2, \ldots, g\}$.  In cell population $pop(t)$ observed at the beginning of day $t$, the initial frequency of $j$-cells (cells with genotype $j$) is denoted as $H_t(j)$, and the population histogram  $H_t=\left[ H_t(1), \dots, H_t(g) \right]$ characterizes the state of $pop(t)$. 
On day $t$, the initial population $pop(t)$ goes through three successive phases: (i) deterministic growth, (ii) random mutations, and (iii) random selection of a subsample of fixed size $N$, which then becomes $pop(t+1)$. Hence, all populations $pop(t)$, $t=1,2,\ldots$, have the same fixed (large) size $N$.
\begin{definition}
The set of all possible histograms $\calH =\{H \in \bR^g\}$ are vectors  of length $g$ such that $0 \le H(j) \le 1$ and $\sum_j H(j) = 1$. Please note that $\calH \subset \bR^g$ is compact and convex.
\end{definition}

\subsection{First Phase: Daily Deterministic Growth}
\label{ss:growth}

During the deterministic growth, the number of $j$-cells increases from initial size $N H_t(j)$ to the final size $N H_t(j) F_j$, where $F_j >1$ is a fixed multiplicative \emph{growth factor}.  The growth factor can be computed as $F_j = \exp(\Delta t \times a_j)$, where $\Delta t$ is the duration of deterministic daily growth (assumed fixed in our model) and $a_j$ is the fitness of genotype $j$. We assume that genotypes are ordered by their fitness, i.e., the vector $F = [F_1,  \dots, F_g]$ of growth factors is ordered by increasing fitness, so that $F_1 < \dots <F_g$, and $g$ is called the \emph{dominant genotype}. At the end of the growth phase, $pop(t)$ reaches the size $N \langle F,H_t \rangle$, and the initial histogram $H_t$ becomes $\Phi(H_t)$, where the function $\Phi: \calH \to \calH$ is given by
\begin{equation} \label{Phi}
\Phi_j (H) = F_j H(j) / \langle F,H \rangle
\quad\text{for all}\quad j= 1, \dots, g, \;\; \text{and} \;\; H \in \calH.
\end{equation}

\noindent Here, $\langle\,\cdot\,, \,\cdot\, \rangle$ denotes the standard inner product of two vectors in $\bR^g$.

\subsection{Second Phase: Daily Random Mutations}
\label{ss:mut}

In bacterial evolution, each time a cell splits into two new cells, random mutations between genotypes may occur with very small probabilities. We have studied this type of daily dynamics using stochastic differential equations with Poisson noise in \cite{schoneman_thesis}. However, in this work, focused on the computational aspects of rare event analysis, we simplify the daily population dynamics by assuming that all random mutations occur simultaneously at the end of each daily deterministic growth period.

In the model considered here, during the day $t$ mutation phase, all individual cells may randomly mutate, independently of each other. Each $j$-cell has a very small probability $m \le 10^{-6}$ of mutating, and the conditional (given that a mutation occurs) probability of a $j$-cell mutating into a $k$-cell is given by a fixed number $0 \le q_{j,k} \le 1$ for $k \neq j$ and $q_{j,j}=0$. For each $j$, we assume that, given $H_t$, the random number of mutants emerging among $j$-cells is Poisson distributed with mean $mNF_jH_t(j)$. Thus, the key mutation parameters are
\begin{itemize}
\item
a small \emph{mutation rate} $0 < m \leq 10^{-6}$,
\item
the $g \times g$ matrix $Q =\{q_{j,k}\}$ of genotype transition probabilities with $\sum_k q_{j,k} =1$ and $q_{j,j} =0$.
\end{itemize}

Denote $\Rt \in \bR^g \times \bR^g$ a random matrix with $R_{j,k}$ describing the number of $j$-cells mutating into $k$-cells on day $t$. The main diagonal of the mutation matrix $\Rt$ is equal to 0, and its conditional mean, given $H_t=H$, verifies
\begin{equation} \label{rho(H)}
\bE(\Rt/N | H_t=H)= \rho(H) \;\;\text{with coefficients}\;\; \rho(H)_{j,k}=m q_{j,k} F_j H(j),
\end{equation}

\noindent where $\Rt/N$ denotes a normalized random mutation matrix with $N$ being the number of cells in the population at the beginning of each day. 

\subsection{Third Phase: Daily Random Selection}
\label{ss:multinom}

After deterministic growth and random mutations, the population $pop(t)$ becomes a much larger population $\mathit{POP}(t)$ of size $N \langle F,H_t\rangle$. Typically, in \textit{E. coli} laboratory experiments, $\operatorname{size}(\mathit{POP}(t)) \gg 200N$. At the end of the day $t$, the \emph{daily selection phase} is implemented by selecting from $\mathit{POP}(t)$ a \emph{random sample} of fixed size $N$ and denoting this sample as $pop(t+1)$. On day $t+1$, $pop(t+1)$, in turn, undergoes growth, mutation, and end-of-day random selection of $N$ cells. For \textit{E. coli} experiments daily random selections are often implemented by dilution of $\mathit{POP}(t)$. Next, we introduce some notation and discuss  a mathematical model for dilution.

\begin{definition} \label{Nrational}
Matrix $A$ is called \emph{$N$-rational} if for a positive integer $N$ all the coefficients of $N A$ are non-negative integers.  
\end{definition}

Given $H_t=H$ and $\Rt/N= r$, the random population histogram $J_t$ of $\mathit{POP}(t)$ is given by
\begin{equation} \label{Psi}
J_t(j) = \Psi_j(H, r) = \frac{1}{\langle F,H \rangle} \left( H(j) F_j - \sum_{k} r_{j,k} + \sum_k r_{k,j} \right).
\end{equation}

\noindent Here, $N \langle F,H \rangle \sum_{k} r_{j,k}$ and $N \langle F,H \rangle \sum_k r_{k,j}$ is the total number of cells mutating ``out'' of and ``into'' the genotype $j$, respectively. The random histograms $H_t$ and the normalized random mutation matrices $r= \Rt/N$ are \emph{N-rational} by construction.

Since $pop(t+1)$ is a random subsample of size $N$ extracted from  $\mathit{POP}(t)$, the conditional distribution of $NH_{t+1}$ given the histogram $J_t = J$ of $\mathit{POP}(t)$ is a multinomial distribution $\mu_{N,J}$ defined by
\[
\mu_{N,J}(v)=N! \prod_{j=1 \dots g} \frac{(J(j))^{v(j)}} {v(j)!}
\]

\noindent for any vector $v = [v(1), \dots, v(g)]$ of non negative integers with $\sum_j v(j) = N$. More precisely, for any $N$-rational histogram $G$, one has
\[
P(H_{t+1} = G \mid J_t = J) = P(N H_{t+1} = N G \mid J_t = J) = \mu_{N,J}(N G).
\]

\subsection{Markov Chain of Population Histograms: Mean Trajectories}
\label{meanpath}

\begin{definition}
$H$ is an interior histogram if $H \in \calH$ and $H(j) > 0$ for all $j=1,\ldots,g$.
\end{definition}

\begin{definition}
The distance between two histograms $H,H' \in \calH$ is given by
\[
\|H - H'\| = \sup_{j=1,\ldots,g} |H(j)-H'(j)|.
\]
\end{definition}

The preceding succession of three daily phases (growth, mutations, and selection) defines a stochastic process $H_t$ as a discrete Markov chain $Pr(H_{t+1} = H' \mid H_t = H)$ with trajectories taking values in the compact convex set $\calH \subset \bR^g$ of population histograms, where  $t=1,2,\ldots$, indexes successive days. Using the mathematical formalism for the three daily stages discussed above, one can compute transition probabilities $Pr(H_{t+1}=H' \mid H_t = H)$ for any histograms $H', H \in \calH$. Since at the beginning of each day, we consider populations of fixed size, $N$, all trajectories consist of $N$-rational $g$-dimensional vectors and, thus, the Markov chain $H_t$ has a finite number of states. The path space $\Omega_T$ of this Markov chain for $1\leq t \leq T$ is the set of all sequences  $\bh= \{H_1,\ldots, H_T\}$ of $N$-rational population histograms, endowed with the distance  
$\|\bh-\bh’\|= \max_{t=1,\dots,T} \|H_t - H’_t\|$.

The \emph{mean trajectory} $\MTR_t=\bE(H_t \,|\, H_1=H)$ is recursively computed as~\cite{azencott2018rare}
\begin{equation}\label{e:MTR}
\MTR_{t+1}(H) = f(\MTR_t(H)),
\end{equation}

\noindent where the function $f$ is the \emph{one-step conditional mean} $f(H) = \bE (H_{t+1} | H_t = H)$ given by
\begin{equation} \label{e:cond.mean}
f_j(H) = \frac{1}{\langle F,H\rangle}
\left(F_jH(j) - m \sum_k q_{j,k} F_j H(j) + m \sum_k q_{k,j} F_k H(k)\right),
\quad j=1 \dots g.
\end{equation}

\noindent For a fixed population size $N$, the mean trajectory $\mathbf{\MTR}$ verifies  $\lim_{t\to\infty} \MTR_t= \hatH$, where the histogram $\hatH$ is given by $\hatH(g) = 1$ and $\hatH(j) = 0$ for all $j<g$. Thus, the limit mean histogram $\hatH$ exhibits a fixation of the genotype $g$ having the highest daily growth factor $F_g$.

The key parameters of this Markov chain model are a large population size $N\ge 10^5$ and a small mutation rate $m \le 10^{-6}$. The biological context is characterized by the fixed set of evolution parameters $\{g,F,Q\}$, where
\begin{itemize}
\item
$g$ is the number of genotypes,
\item
$F = \left[F_1, \ldots, F_g\right]$ is the vector of ordered daily growth factors, and
\item
$Q$ is a $g \times g$ matrix with $q_{j,k}$ being the conditional (given that a mutation occurs) probability that a mutant $j$-cell becomes a $k$-cell.
\end{itemize}

For instance, in the \textit{E. coli} experiments \cite{heger,azencoop}, parameters  of the Markov chain $H_t$ can typically be selected in the following range
$10^5 \leq N \leq 10^8$, $10^{-9} \leq m \leq 10^{-6}$,
$5 \leq g \leq 10$, and
$200 \leq F_1 < \dots < F_g \leq 500$.  
For many genetic experiments, the conditional probability $q_{j,k}$ that a mutant $j$-cell becomes a $k$-cell cannot be directly estimated from data. Hence, for simplicity, we assume uniform genotype transition probabilities $q_{j,k} = 1/(g-1)$ for all $k\neq j$ with $q_{j,j} =0$.

\section{Rare Events and Large Deviation Asymptotics}
\label{sec3}

In a previous paper~\cite{azencott2018rare}, we have rigorously studied the large deviation asymptotics for the Markov chain of population histograms. Those results are applicable as soon as  $N \ge 10^5$ and $m \le 10^{-6}$, and we now recall the main results proved in \cite{azencott2018rare}. Fix any $T>0$. Any random events of interest concerning trajectories of the Markov chain can be defined  as a subset $\mathcal{E} \subset \Omega_T$.  For increasing $N \ge 10^5$ and fixed initial histogram $H_1=H$, the random paths $\mathbf{H}$ starting at $H_1= H$ have a probability distribution $P_N$ on $\Omega_T$ which becomes highly concentrated around the mean  trajectory $\mathbf{\MTR}$ starting at $H$. Indeed, any closed subset $\mathcal{E}$ of $\Omega_T$ such that $\mathbf{\MTR} \notin \mathcal{E}$ has probability $P_N(\mathcal{E})$ vanishing at exponential speed as $N\to\infty$. For large $N$, $\mathcal{E}$ becomes a rare event and we introduce a more precise definition.

\begin{definition} \label{log-equivalent}
For fixed initial histogram $H_1= H$ and time $T>1$ we call $\mathcal{E} \subset \Omega_T$ a rare event if $\frac{1}{N}\log(P_N(\mathcal{E})) \to - \Lambda(\mathcal{E})$ as $ N \to \infty$. Here, $\Lambda(\mathcal{E}) = \min_{\bh \in \mathcal{E}} \lambda(\bh)$, where $\lambda(\bh)$ is the rate functional (or the cost of the path $\bh$) defined in~\eqref{e:firstlambda}. As a short-hand notation we then write $P_N(\mathcal{E})$ is log-equivalent to $\exp(- N\Lambda(\mathcal{E}))$.
\end{definition}

\subsection{Most Likely Histogram Trajectory Linking two Population Histograms}

\begin{definition}
A path $\bh= \{H_1, \ldots, H_T\} \in \Omega_T $  is called an \emph{interior path} if all histograms  $H_t$ are interior histograms, i.e., they verify  $H_t(j) > 0$ for $j=1, \ldots,g$.
\end{definition}

Denote $\mathcal{T}_N \subset \Omega_T$ the \emph{thin tube} of all paths in $\Omega_T$ lying within distance $1/N$ of  $\bh$. Then (see \cite{azencott2018rare}), as $N \to \infty$  the probabilities  $P_N(\mathcal{T}_N)$ are \emph{log-equivalent} to  $\exp(-N\lambda(\bh))$, where the \emph{rate function} $\lambda(\bh ) \geq 0$ is given by
\begin{equation} \label{e:firstlambda}
\lambda(\bh) = \sum_{t=1}^{T-1} C(H_t, H_{t+1}).
\end{equation}

\noindent Here, $C(H,G)\geq 0$ is a computable one-step cost function (see \eqref{e:C(H,G)} below).

The \emph{most likely path} $\MLP({H,G})=\{L_1 = H, L_2, \ldots, L_T = G\}$ connecting  histograms $H$ and $G$ in $T$ steps is  then determined by minimizing $\lambda(\bh)$ over all paths  $\bh= [ H_1, \ldots, H_T ]$ such that $H_1=H$ and $H_T= G$, i.e.
\[
\MLP({H,G}) = \bh^\opt = \argmin_{\bh \in \Omega_T^+(H,G)} \lambda(\bh) ,
\]

\noindent where 
\begin{equation}\label{omegaplus}
\Omega_T^+(H,G) = \{\bh \in \Omega_T : H_1 = H \;\text{and}\; H_T = G\}
\end{equation}
 
\noindent is a restricted space of paths.

As shown in~\cite{azencott2018rare}, any most likely path  $\bh^\opt=[H^\opt_1,\ldots, H^\opt_T ]$ must verify a second order recursive equation expressing explicitly $H^\opt_t$ in terms of $H^\opt_{t+1}, H^\opt_{t+2}$. This paper explores efficient strategies for computing the path $\bh^\opt$ since such numerical computations are particularly challenging if the number of genotypes $g$ exceeds 4.

\subsection{The One-Step Cost Function $C(H,G)$}

On each day $t$, the initial random histogram $H_t$, the normalized matrices $r_t= R_t/N$ of random mutations, and the histogram $J_t$ of $\mathit{POP}_t$ after mutations, must verify the deterministic relation $J_t = \Psi(H_t, r_t)$ in~\eqref{Psi}. Since the number $N\mu_t(j)$ of mutants among the $j$-cells of $\mathit{POP}_t$ is inferior to the size $N F_j H_t(j)$ of the $j$-cells colony, $r_t$ must belong to the set $\calM(H_t)$ of $g\times g$ matrices with entries $r_{j,k}$ defined by
\begin{equation} \label{K(H)}
\calM(H) = \left\{
r : r_{j,k} \geq 0; \;\;  r_{j,j} = 0; \;\; \text{and} \;\; \sum_{k=1}^g r_{j,k} \leq  H(j) F_j \;\;\text{for all}\; j
\right\}.
\end{equation}

\noindent Thus, $\mu_t(j)$ have independent slightly truncated Poisson distributions with means $m F_j H_t (j)$.

Fix any interior histogram $H$ and any matrix $r \in \calM(H)$. 
Denote $J= \Psi(H,r)$. As a short-hand notation we use $H_t \approx H$,
$r_t = R_t/N \approx r$, whenever the $\sup$-norms $\|H_t – H\|$ and 
$\|r_t – r\|$ are bounded by $c/N$ for some fixed constant $c > 0$. One 
then has a similar bound for $\|J_t – J\|$, for which we also write 
$J_t \approx J$. The rate function controlling large deviations from 
the mean for products of truncated Poisson distributions is determined 
in~\cite{azencott2018rare} to show that for any interior histogram $H$, 
any matrix $r \in \calM(H)$, and for $N\to \infty$ one has
\begin{align}
& Pr(r_t \approx r \;|\; H_t \approx H) \;\; \text{is log-equivalent to} \; \exp(- N \mut(r,H)), \;\; \text{where}  \label{raremutation} \\
& \mut(r,H)=\sum_{j,k}  r_{j,k} \log r_{j,k}  - r_{j,k} \left[ 1 + \log(m q_{j,k}F_jH(j)) \right] + m q_{j,k}F_jH(j). \label{mut(r,H)}
\end{align}

Given $J_t$, the conditional distribution of  $N H_{t+1}$ is a multinomial distribution with mean $N J_t$, and its large deviations from the mean are controlled by an explicit rate function. Indeed, for any two interior histograms $(G, J)$, and for $N\to \infty$,  one has~\cite{azencott2018rare}
\begin{equation} \label{raremultinomial}
Pr(H_{t+1} \approx G \mid J_t  \approx J) \;\;\text{is log-equivalent to}\;\exp( - N KL(G,J) ),
\end{equation}

\noindent where the \emph{Kullback--Leibler divergence} $KL(G,J) \geq 0$ is finite, and given by
\begin{equation}\label{KL}
KL(G,J) = \sum_j G(j) \log \frac{G(j)}{J(j)}
\end{equation}

Since $J= \Psi(H,r)$, one has
\[
Pr(\{H_{t+1} \approx G \} \;\text{and} \; \{r_t \approx r\}\,|\, H_t \approx H) = Pr(H_{t+1} \approx G \;|\; J_t \approx J) P(r_t \approx r \,|\, H_t \approx H)
\]

\noindent so that, due to~\eqref{raremultinomial} and~\eqref{raremutation},
\[
Pr(\{H_{t+1} \approx G \} \;\text{and} \; \{r_t \approx r\}\,|\,H_t \approx H) \;\;\text{is log-equivalent to}\; \exp( - N v(H,r,G)),
\]

\noindent where $v(H,r,G) = \mut(H,r) + KL(G,\Psi(H,r))$.

The approximate one-step transition probability $Pr(\{H_{t+1} \approx G \} \,|\,H_t \approx H)$ is then log-equivalent to the sum over all $r \in \calM(H)$ of the vanishing exponentials   $\exp(- N v(H,r,G))$. This sum is actually log-equivalent to the largest of all these terms, namely $\exp(- N C(H,G))$ where the \emph{one-step cost function} $C(H,G) \geq 0$ is defined by
\begin{equation} \label{e:C(H,G)}
C(H,G)=\min_{r \in \calM(H)}  v(H,r,G) = \min_{r \in \calM(H)} \left[ \mut(r,H) + KL(G, \Psi(H,r)) \right].
\end{equation}

Therefore, for arbitrary interior histograms $H,G$ and $N\to\infty$, we have the asymptotic result 
\begin{equation}\label{e:one.step.jump}
Pr(\{H_{t+1} \approx G \} \,|\, H_t \approx H) \;\;\text{is log-equivalent to} \; \exp(-N C(H,G))).
\end{equation}

\noindent The cost $C(H,G)\geq 0$ can only take the value $0$ when $G=f(H)=\bE(H_{t+1}\;|\;H_t=H)$, with $f(H)$ given by~\eqref{e:cond.mean}. Hence, for $H$ fixed and large $N$, the Markov transition kernel is concentrated at exponential speed (in $N$) on a fixed  small neighborhood of $f(H)$. When $H,G$ are \emph{interior} histograms, i.e. when $\min(H)>0$ and $\min(G)>0$, the cost $C(H,G)$ is a finite smooth function of $H$ and $G$. Moreover, for a small enough mutation rate $m \le 10^{-6}$, the cost $C(H,G)$ has an explicit expansion in $m$ given by
\begin{equation} \label{e:costformula}
C(H,G) = KL(G,\Phi(H)) + m \sum_{j,k} F_j H(j) q_{j,k} [1-U_k/U_j] + O(m^2),
\end{equation}

\noindent where $U_j=\exp \left(\frac{G_j}{F_jH(j)}\right)$ and 
$\Phi \in \bR^g$ with 
$\Phi(j) = \frac{F_jH(j)}{\langle F,H\rangle}$.

\subsection{Large Deviation Asymptotics in Path Space}

The path space $\Omega_T$ of the Markov chain $H_t$ is endowed with a natural distance (see \secref{meanpath}). For any path $\bh=\left[H_1, \dots, H_T\right]$ in $\Omega_T$ and any small  $\epsilon>0$ define the closed ``tube'' of paths $\mathcal{T}(\bh,\epsilon)$ centered at the path $\bh$ by
\begin{equation} \label{tube}
\mathcal{T}=\{\bh'\in\Omega_T\;\;\text{such that}\;\|\bh'-\bh\| \leq \epsilon\}.
\end{equation}

\noindent We say that $\mathcal{T}$ is an \emph{interior tube} if all $\bh'\in\mathcal{T}$ are interior paths. We extend the one-step cost $C(H,G)$ to a \emph{multi-steps} cost functional $\lambda(\bh)\geq 0$ defined for any $\bh \in \Omega_T$ by
\begin{equation} \label{costintegral}
\lambda(\bh) = \sum_{t=1}^{T-1}  C(H_t, H_{t+1}).
\end{equation}

The rate function $\lambda(\bh)$ defined for paths $\bh \in \Omega_T$ can be formally extended to a rate functional $\Lambda(A)\geq0$ defined for all subsets $A \subset \Omega_T$ by
\begin{equation} \label{e:LAMBDA}
\Lambda(A) = \inf_{\bh \in A}  \; \lambda(\bh).
\end{equation}

\noindent The cost function $C(H,G)$ is continuous in $H$ and $G$ on the set of interior histograms. This implies that when $\bh$ is any fixed interior path, then
$\Lambda(\mathcal{T}(\bh,\epsilon)) \to \lambda(\bh)$ as $\epsilon \to 0$.
As $N \to \infty$, we proved in~\cite{azencott2018rare} that $\lambda(\bh)$ is the large deviations \emph{rate functional} controlling probabilities of rare events for our Markov chain. This result is formulated as follows.

\begin{theorem}\label{probatube}
Fix an interior path  $\bh=[H_1, ..., H_T] \in \Omega_T$, and  an interior tube of paths $\mathcal{T}(\bh, \epsilon)$. Denote  $\bhh=[h_1, ..., h_T]$ a random path of population histograms staring at $h_1=H_1$. Then, as $N \to \infty$, the probabilities $P_N(\bhh \in \mathcal{T})$ are \emph{log-equivalent} to  $\exp(-N \Lambda(\mathcal{T}))$, where the set functional $\Lambda$ is defined by \eqref{e:LAMBDA}.
More precisely, the ``thin'' tubes of paths $\mathcal{T}(\bh,1/N)$ have probabilities $P_N(\bhh \in \mathcal{T})$ which are \emph{log-equivalent} to $\exp(-N\lambda(\bh))$ as $N\to\infty$.
\end{theorem}

\noindent\textit{Proof:} See~\cite{azencott2018rare}.\vspace{0.1cm}

The non-negative rate functional $\lambda(\bh)$ can only reach the value $0$ when $\bh$ is  the mean trajectory $\mathbf{\MTR}$ starting at $H_1$ (see  \eqref{e:MTR}). Thus, mean trajectories are also called \emph{zero-cost paths}. When the initial condition of the Markov chain is fixed at $H_1$ and as $N \to \infty$, the probabilities $P_N(\bh)$ vanish exponentially fast if $\bh$ is not the zero-cost path $\mathbf{\MTR}$ starting at $H_1$ but tends to 1 exponentially fast if $\bh=\mathbf{\MTR}$.

Consider fixed histograms $H$ and $G$ and recall the definition of the space of restricted paths $\Omega_T^+(H,G)$ in~\eqref{omegaplus}. Consider paths $\bh \in \Omega_T^+(H,G)$ and the corresponding set of all tubes $\mathcal{T}(\bh,1/N)$. Then, the most likely tube $\mathcal{T}(\bh^\opt,1/N)$ is centered around the path $\bh^\opt$ which solves the minimization problem
\begin{equation}\label{minlambda}
    \bh^\opt = \argmin_{\bh \in \Omega_T^+(H,G)} \lambda(\bh),
\end{equation}
\noindent where $\lambda(\bh)$ is given by \eqref{costintegral}.
For interior histograms $H$ and $G$, this minimization problem always has at least one solution $\bh^\opt$ which of course depends on the number of steps $T \geq1$. Any such minimizing path is a \emph{most likely path} linking $H$ to $G$ in $T$ steps. Indeed, for the Markov chain of population histograms, and for large $N$, following the thin tube $\mathcal{T}(\bh^\opt,1/N)$ in path space maximizes the probability of reaching histogram $G$ at time $T$ given $H_1 = H$. In fact, for large $N$, both $Pr(H_T=G\;|\;H_1=H)$ and $Pr(\mathbf{H} \in \mathcal{T}(\bh^\opt,1/N)$ are \emph{log-equivalent} to $\exp(- N \lambda(\bh^\opt))$. Note that the optimal rate $\lambda(\bh^\opt)$ also depends on $T$.

In long-term studies of bacterial evolution, one may know that the  population evolved from a remote past ancestor histogram $H$ to a currently observed histogram $G$, without knowing the precise time duration between these two observations. This leads to the problem of finding both the unknown time $T$ and the \emph{most likely path} linking $H$ to $G$ in $T$ steps. In our numerical benchmark studies below, we implemented this type of computation by numerically minimizing in $T$ the optimal rate $\lambda(\bh^\opt)$. Our computational strategies outlined in this paper yield both an optimal number of steps $T^\opt$ and a  most likely path $\bh^\opt(T^\opt) \in \Omega_{T^\opt}^+(H,G)$. The uniqueness of $T^\opt$ and $\bh^\opt(T^\opt)$ has not been proved, but our numerical computations of $\bh^\opt(T^\opt)$ for many random  pairs $(H,G)$ indicate that  uniqueness is likely to hold for almost all interior histograms $H$ and $G$.

\subsection{Analogies between Most Likely Population Paths and Riemannian Geodesics} 
\label{analogies}

For many classes of continuous time Markov processes living in $\bR^k$, large deviation rate functions analogous to $\lambda$ have been explicitly determined. For instance, for stochastic differential equations ({\bf SDE}s) with noise multiplied by a small parameter $\epsilon$, denote $p_{\epsilon}(f)$ the probability that a random SDE path remains within a thin tube around the smooth path $f_t \in \bR^k$ for all $t\leq T$. Then (see~\cite{azenLDT2012, ventcell}), $p_{\epsilon}(f)$ is roughly equivalent to $\exp(-\lambda(f)/\epsilon^2 )$, with a rate function $\lambda(f) \geq 0$ given by
\begin{equation} 
\label{generalcost}
\lambda(f) = \int_{\left[0,T\right] } \mathit{Cost}(f_t, f'_t)\,\mathrm{d}t \;\; \text{when}\; f'_t \in L_2(0,T),
\end{equation}

\noindent where $\mathit{Cost}(u,v) \geq 0$ is an explicit smooth function of $u,v \in \bR^k$. Given $x,y \in \bR^k$, the most likely SDE path $f$ such that $f_0 = x$ and $f_T=y$, is determined by minimizing $\lambda(f)$ over all paths $f$ subject to $f_0 = x$ and $f_T=y$. All minimizing paths $f^\opt$  must  verify  $\partial_f \lambda(f) = 0$, where  $\partial_f \lambda$ is the differential of $\lambda(f)$. This yields a non-linear, second-order differential equation verified by $f^\opt$ for all $t$, to be solved  under the constraints $f^\opt_0=x$, $f^\opt_T=y$.

On any Riemannian manifold $\mathcal{S}$, all \emph{geodesics} $f_t \in S$ must verify an analogous second-order ordinary differential equation ({\bf ODE}). Indeed, a \emph{geodesic} linking $f_0 =x$ to $f_T=y \neq x$ in $\mathcal{S}$ is a minimizing path for the \emph{kinetic energy} $\mathit{Kin}(f)$ defined by
\begin{equation} 
\label{kinetic_energy}
\mathit{Kin}(f) = \frac{1}{2}
\int_{\left[ 0,T \right]}
\left\langle f'_t, \gamma_{f_t} f'_t \right\rangle
\, \mathrm{d}t \;\; \text{when}\;  f'_t \in L_2(0,T),
\end{equation}
where the quadratic form $\gamma_z$ is the Riemann metric at $z \in \mathcal{S}$. Note that if one sets
\[
\mathit{Cost}(f_t, f'_t)  = \frac{1}{2} 
\left\langle f'_t, \gamma_{f_t} f'_t \right\rangle
\]

\noindent then $\mathit{Kin}(f)$ has the same form as $\lambda(f)$ given in \eqref{generalcost}.

Minimizing the kinetic energy $\mathit{Kin}(f)$ as well as minimizing the SDE rate function $\lambda(f)$ over all smooth paths $f_t$ such that $f_0=x$ and $f_T=y$ lead to solving two similar second-order ODEs with endpoint constraints. Similarly, minimizing our rate functional $\lambda(\bh)$ over all discrete-time paths $\bh \in\Omega_T^+(H,G)$ leads (see the next section) to solving in reverse time a recursive equation of order two, roughly similar to the second-order ODEs for continuous-time problems discussed above. Thus, in the present paper, the most likely path $\bh^\opt \in\Omega_T^+(H,G)$ connecting two population histograms $H$ and $G$ will also be called a \emph{geodesic} from $H$ to  $G$.

\subsection{Reverse Time  Computation of Most Likely Paths}
\label{s:reversegeo}

Any most likely path $\bh^\opt \in \Omega_T^+(H,G)$ must solve the cost minimization problem~\eqref{minlambda}. As proved in~\cite{azencott2018rare} for small mutation rate $m$, any such minimizing path $\bh^\opt \in  \Omega_T^+(H,G)$ is \emph{fully determined} by its last two histograms $H_T^\opt= G$ and $H_{T-1}^\opt$. Indeed, $\bh^\opt$ must verify \emph{the reverse time recursive equation}
\begin{equation} \label{e:reverse}
H_t^\opt = \chi(H_{t+1}^\opt, H_{t+2}^\opt) \;\;\text{for all} \; \; 1 \leq t \leq T-2,
\end{equation}

\noindent where $\chi(y,z)$ is an explicit histogram-valued smooth function defined for all interior histograms $(y,z)$. For small mutation rate $m \le 10^{-6}$, the first-order Taylor expansion of $\chi(y,z)$ was computed in \cite{azencott2018rare} as follows:
%
%
\begin{equation} \label{e:chi}
\chi(y,z) \simeq x + O(m^2) \;\;\text{ with} \; x_j = Y_j (1+ m w_j ) \;\;\text{for} \; j= 1, \dots, g,
\end{equation}

\noindent where, given the histograms $y$ and $z$, one computes the interior histogram $Y$ and the vector $w\in \bR^g$  by the following explicit formulas (where $j,k =1 \dots g$)
\begin{align*}
& X_j = \frac{y_j}{F_j}  \exp \left( \frac{F_j}{\langle F, y \rangle} - \frac{z_j}{y_j} \right) >0, \qquad
Y_j = \frac{X_j}{\sum_k  X_k},\\
&  e_{j,k} = \exp \left[ -y_j/(F_j Y_j) + y_k/(F_k Y_k)\right], \\
& \alpha_j =  \sum_k  \left( q_{j,k} e_{j,k} - \frac{F_k X_k}{F_j X_j} q_{k,j} e_{k,j} \right),\\
& f_{j,k} = \exp \left[-z_j/(F_j y_j) + z_k/(F_k y_k)\right],\\
& \beta_j = F_j \sum_k q_{j,k} - \left( F_j + \frac{z_j}{y_j} \right) \sum_k  f_{j,k} q_{j,k} - \frac{z_j}{F_j y_j^2} \sum_k  F_k y_k q_{k,j} f_{k,j},\\
& w = \alpha + \beta - \langle Y , \alpha +\beta \rangle.
\end{align*}

\noindent The reverse recurrence~\eqref{e:reverse} reduces the computation of $\bh^\opt$ to the optimization search with respect to the penultimate histogram
\[
Z^\opt = \argmin\limits_{Z \in \calH^+} \lambda([H_1 \equiv H, H_2, \ldots, H_{T-1} = Z, H_T \equiv G]),
\]

\noindent where $\calH^+ =\{Z: Z \in \calH \text{ and } Z(j) > 0\}$ is the  set of all possible interior histograms. Then, the optimal trajectory is given by $\bh^\opt = [H, H_2, \ldots, H_{T-1}\equiv Z^\opt, H_{T} \equiv G]$, where $H_t$ are computed from~\eqref{e:reverse} for $t=2, \ldots ,T-2$. Note that condition $H_1=H$ is always enforced; consequently, we use the reverse recursive relationship \eqref{e:reverse} only until $t=2$.
We always seek $H_{T-1} = Z^\opt$ among interior histograms so that the path $\bh$ generated by equation~\eqref{e:reverse} is an interior path. Next, we consider this numerical problem, which is similar to reverse geodesic shooting on Riemannian manifolds, and is quite challenging when $g \ge 4$.

We call the $T$-steps cost minimizing path $\bh^\opt \in \Omega_T^+(H,G)$ the \emph{geodesic of length $T$} from $H$ to $G$, and we denote its (minimized) cost $\lambda(\bh^\opt)$ by $\lambda_T^\opt(H,G)$. To fully determine the most likely path when the number $T$ of path steps is allowed to be arbitrary, one needs to compute \[\lambda^\opt(H,G) = \min_{T \geq 1} \lambda^\opt_T(H,G)\] and the optimal number of steps $T^\opt$ such that $\lambda^\opt(H,G) = \lambda^\opt_{T^\opt}(H,G)$. For given interior histograms $H,G$, and mutation rates $m$ small enough, we conjecture that a finite $T^\opt$ always exists, as supported by numerical simulations outlined below.

For any interior geodesic $\bh=[H_1,\ldots,H_T=G]$, all the intermediary histograms $H_t$ are fully determined by both $G$ and the \emph{penultimate histogram} $Z = H_{T-1}$ due to the reverse recurrence~\eqref{e:reverse}. In particular, the initial histogram $H_1$ is then a smooth function of $G$ and $Z$ defined iteratively by
\begin{equation} \label{e:thetaT}
\theta_1(Z,G)=G, \quad 
\theta_2(Z,G)=Z, \quad
\theta_t=\chi(\theta_{t-1},\theta_{t-2})\;\;\text{for}\;\; t\geq 3.
\end{equation}

\noindent Here, $\btheta_T = [\theta_T, \theta_{T-1}, \ldots, \theta_2, \theta_1]$ represents the trajectory $\bh = [H_1,\ldots,H_T=G] $ in reverse time, i.e.
$\theta_T = H_1$, $\theta_1 = G$, 
$\theta_{2} = H_{T-1}$, etc.
Notice that it is very unlikely that the reverse computation of the most likely path using~\eqref{e:thetaT} would amount to $\theta_T = H_1$. Therefore, when searching for the most likely path $\bh^\opt \in \Omega_{T^\opt}^+(H,G)$ we use the reverse formula \eqref{e:thetaT} to compute $\theta_t$ for $3 \le t \le T$ and then ``connect'' path $\btheta_T$ to $H$ using the zero-cost mean path starting at $H$. Then, the resulting  most likely path $\bh^\opt$ is given as a concatenation of two paths---the zero-cost mean path and the path $\btheta_T$. Then, $\bh^\opt$ connects $H$ and $G$ by construction and the cost of $\bh^\opt$ is the cost of $\btheta_T$ plus the cost of the ``connection'' between the mean path and $\btheta_T$. This ``connection'' can occur at an arbitrary time $2 \le \tau \le  T$. The key question for the reverse time geodesic shooting is to find a penultimate histogram $Z=H_{T-1}$ and an integer $\tau \geq 2$ such that $\theta_\tau$ is close to the mean path starting at $H$.

\section{Algorithms for Reverse Time Geodesic Shooting}
\label{s:brute_algo}

As discussed in the previous section, in order to compute numerically the most likely path one needs to perform an optimization search with respect to the penultimate histogram $H_{T-1} = Z$. This optimization problem can be combined with the search for the optimal number of steps, $T^\opt$. First, we consider the problem of computing the optimal cost $\lambda^\opt_{T}(H,G)$ for a fixed number of steps, $T$.
%
%
To this end, we need to fix a set $\calS \subset \calH^+$ of interior penultimate histograms. (We discuss how to select $\calS$ efficiently in \secref{ss:fullbrute} and \secref{ss:p-quantiles}, respectively.) We compute the optimal penultimate histogram by performing an optimization over $Z \in \calS$, i.e.,
\[
Z^\opt = \argmin\limits_{Z \in \calS} \lambda([H_1 \equiv H, H_2, \ldots, H_{T-1} = Z, H_T \equiv G]),
\]

\noindent and an approximate geodesic is given by $\bh^\opt = [H_1 \equiv H, H_2, \ldots, H_{T-1} = Z^\opt, H_T \equiv G]$, where $H_t$ are computed from~\eqref{e:reverse} for $t=2, \ldots ,T-2$. However, we need to account for the fact that one of the histograms $H_k=\chi(H_{k+1}, H_{k+2})$ (for $1< k < T-1$) can be a boundary histogram. In this case, we cannot continue reverse shooting by using~\eqref{e:reverse}. Instead, we connect $H_k$ with a mean trajectory starting at $H$. This approach is discussed in the next section.

\subsection{Computation of Approximate Geodesics}
\label{ss:long geo}

In order to define a boundary histogram,  we first define a small number $\epsilon(N)$. For practical values $m \leq 10^{-6}$ and $N \ge 10^5$, it can be defined as $\epsilon(N) =  50/N$, so that $\epsilon(N) \le 5 \times 10^{-5}$. Next, we define a boundary histogram where at least one of the genotypes is nearly extinct, i.e.,

\begin{definition}
A histogram $H \in \calH$ is a \emph {boundary histogram} if $\min_{j= 1 \dots g} H(j) \le \epsilon(N)$.
\end{definition}

Next, in order to combine the optimization of the cost function with respect to the penultimate histogram  $Z \in \calS$ and the number of steps, $T$, we define the maximum possible number of steps in the optimal trajectory, $T_{\max}$. In practice, one can safely take $T_{\max} = 4 g$ since rare event geodesics with strictly positive cost have a relatively small number of steps (see \secref{sec8}).

Then, for each $Z \in \calS$ we perform the reverse shooting using the following two steps:\\
\textbf{Step 1.}
Compute recursively the \emph{open ended} reverse time trajectory $\btheta$ by the reverse time recursion~\eqref{e:reverse}
\[
\theta_1=G, \; \theta_2= Z, \;\text{and}\;\theta_{t} = \chi(\theta_{t-1}, \theta_{t-2}) \;\text{for}\; t \geq 3.
\]
The iterative computation of $\theta_t$ is stopped at time $\tau$ if either $\tau = T_{\max} - 1$ or if $\theta_{\tau}$ is a boundary histogram. \\
\textbf{Step 2a.}
If $\tau = T_{\max} - 1$, then $\theta_\tau$ is an interior histogram and we ``connect'' the reverse path to $H$ by assigning $\theta_{T_{\max}} = H$. \\
\textbf{Step 2b.} If $\tau < T_{\max} - 1$, then $\theta_\tau$ is a boundary histogram and we ``connect'' the reverse path using the mean trajectory starting at $H$. In particular, the mean trajectory $\mathbf{\MTR} = [H, M_2, \ldots, M_{T_{max} - \tau}]$ of length $T_{\max} - \tau$ is computed using \eqref{e:MTR}. Notice that the cost of the mean trajectory is always zero. Next, we perform a search for two indexes $(u,v)$ such that cost $\lambda(M_u, \theta_v)$ is minimal among all $(u,v)$ such that $1 \le u \le T_{max} - \tau$ and $1 \le v \le \tau$. Notice that this search is fast since $T_{\max}$ is not very large. Then, the trajectory connecting $H$ and $G$ is given by $\bh = [H, M_2, \ldots M_u, \theta_v, \ldots \theta_2, G]$ with the cost given by~\eqref{costintegral}. By construction, the length of this trajectory is less or equal to $T_{\max}$ and the penultimate histogram is $\theta_2 = Z$.

Since $\calS$ is finite, optimization search over all $Z \in \calS$ yields an optimal penultimate histogram  $Z^\opt \in \calS$ 
and the optimal trajectory $\bh^\opt=[H, H_2, \ldots Z^\opt \equiv H_{{T^\opt}-1}, G \equiv H_{T^\opt}]$
with $T^\opt \le T_{max}$ by construction. Of course, the optimal trajectory $\bh^\opt$ and the number of steps $T^\opt$ can potentially depend on the choice of the set $\calS$. Thus, it might be necessary to consider very large sets of penultimate histograms
in order to find a global minimizing trajectory connecting $H$ and $G$. Therefore, we discuss heuristic arguments for constructing 
$\calS$ in order to reduce the number of penultimate histograms and the corresponding computational cost.

\subsection{Brute-Force Search}
\label{ss:fullbrute}

One obvious choice would be to take $\calS = \calH^+$, the set of all interior histograms. However, this choice is computationally intractable, and also not necessary. Instead, one can attempt to use a much coarser discretization of $\calH^+$. Thus, we introduce a parameter $\zeta$ with
$100 \leq \zeta \leq 500$ and denote $\calH^+_\zeta$ the set of all interior $\zeta$-rational histograms. Set $\calH^+_\zeta$ provides discretization of $\calH^+$ with mesh size $1/\zeta$ and we can set $\calS = \calH^+_\zeta$. Note that in this case $\operatorname{card}(\calS) \simeq \zeta^{g-1}$. The number of penultimate histograms in $\calS$  for different values of $g$ and $\zeta$ are listed in \tabref{tableCardH}. One can see that it is relatively easy to implement a straightforward search over all $Z \in \calS$ for $g \le 4$ genotypes. However, even for a coarse discretization with $\zeta=100$, the direct approach becomes computationally intractable (even with a parallelization on 20 nodes) for $g > 5$.
\begin{table}[h] 
\begin{center}
\begin{tabular}{lrrrrrr} 
	\toprule
	$(g, \zeta)$ & $(3,500)$ & $(4,200)$ & $(5,100)$ & $(6,100)$ & $(7,100)$ & $(8,100)$\\
	\midrule
	$card(\calS)$ &  $1.3\times 10^5$ & $1.4\times 10^6$ & $\simeq 10^8$ & $\simeq 10^{10}$ & $\simeq 10^{12}$ & $\simeq 10^{14}$ \\
	\bottomrule
\end{tabular}
\caption{$card(\calS)$ for different choices of the number of genotypes, $g$,  and  the discretization level, $\zeta$.}
\label{tableCardH}
\end{center}
\end{table}

One can attempt to develop an iterative refinement strategy starting with a coarse discretization of e.g. $\zeta_1=20$. Optimization over the set of all penultimate histograms in $\calS_1 = \calH^+_{\zeta_1}$ will result in an optimal  path and optimal penultimate histogram $Z_1^\opt$. However, this path is unlikely to be the global minimizer due to a coarse discretization of $\calH^+$. Then, it is possible to select $\calS_2$ as a small neighborhood around $Z_1^\opt$ with a finer discretization $\zeta_2 > \zeta_1$. This will result in an optimal penultimate histogram $Z_2^\opt$, which can be refined further. This yields an iterative approach for computing the most likely path and the optimal penultimate histogram $Z^\opt$. However, this iterative approach is still exceedingly expensive for $g \geq 6$. Therefore, we discuss a different approach for selecting the set $\calS$ next.

\subsection{$p$-Quantile Approach for Constructing $\calS$} \label{ss:p-quantiles}

In the case of three genotypes, $g=3$, an extensive computation of geodesics is very fast for $\calS= \calH^+_\zeta$ with $\zeta=500$.  Thus, for $g=3$ we performed extensive numerical investigation and computed many most likely paths connecting pairs $H$ and $G$ selected randomly from the set $\calH^+$. We conjecture that for $g=3$, most likely paths $\bh^\opt \in \Omega^+(H,G)$ have penultimate histograms $Z^\opt$ with a fairly low last step cost $C(Z^\opt,G)$.  This suggests that the set $\calS$ of  penultimate histograms can be selected to minimize the last step cost $C(Z,G)$.
This can be implemented by explicitly computing the derivative of the last step cost and selecting a set $\calS$ with a fairly low Euclidian norm $\|\partial_Z \, C(Z,G)\|$ (see \apxref{s:gradient}). 

Instead of utilizing the derivative of the last step cost, we have developed a more efficient \emph{$p$-quantile} approach for selecting $\calS$. This approach is based on the same conjecture of low last step costs for optimal geodesics.  Consider the set $\calH^+_\zeta$ of interior histograms which discretizes the set of all interior histograms, $\calH^+$ with step $1/\zeta$. Next, compute one-step costs $C(Z,G)$ for all $Z \in \calH^+_\zeta$ and denote $\beta_p$ the $p$-quantile of such one-step costs. Then, the set $\calS\equiv\calS(p)$ of penultimate histograms can be chosen as
\begin{equation}
\label{e:PEN(p)}
\calS(p) = \{Z \in \calH^+_\zeta \;|\; C(Z,G) < \beta_p\}. 
\end{equation}

\noindent For $\zeta=100$, the cardinality of $\calS(p)$ is roughly $p \times 10^{2g-2}$. The choice $\calS=\calS(p)$ always reduces the computing time for the brute-force approach discussed in the previous section by a factor of approximately $p$. Numerical results are presented in \secref{s:gseveneight}.

\subsection{Numerical Results}
\label{sec8}

In this section, we describe numerical results designed to compare the two approaches for constructing the set of penultimate histograms discussed in \secref{ss:fullbrute} and \secref{ss:p-quantiles}, respectively. The main goal is to demonstrate the computational efficiency of the $p$-quantile approach. We also discuss the biological consequences of the computed most likely paths. The parameters of our model are estimated from laboratory experiments on the genetic evolution of \textit{E.coli} bacteria \cite{cooper1,cooper2} and are presented in \apxref{ap:params}. 

All computations in this paper were carried out on the Opuntia multi-node cluster available at the University of Houston. Each of these nodes is equipped with 20 core CPUs (Intel Xeon E5-2680v2 2.8 GHz). Each computational task was divided into 20 approximately equal parts, and all computational parts were carried out in parallel. On each node, computational performance is increased by exploiting (hyper-)threading via MATLAB's parallel computing toolbox.

\subsubsection{Numerical Results for $g=3,4,5$ Genotypes}
\label{ss:parallel.brute}

Runtimes for simulations with both approaches for constructing the set $\calS$ are reasonable  for $g=3,4,5$ genotypes and $\zeta=100$. Thus, we first perform simulations in this parameter regime to compare the approaches presented in 
\secref{ss:fullbrute} and \secref{ss:p-quantiles}, respectively. We selected $10,000$, $5,000$, and $240$ random pairs $(H,G)$ of terminal histograms for $g=3,4$, and $5$, respectively. For each pair $(H,G)$ we then used $\zeta=100$ and performed the search for the optimal trajectory with $\calS = \calH^+_\zeta$.

\begin{table}
\caption{Mean most likely path length computed for $g=3,4,5$.}
\label{table:geolengths}
\centering
\begin{tabular}{lrr}
\toprule
$g$ & \bf \# geodesics & mean length \\
\midrule
$g = 3$ & 10,000 & $5.02 \pm 0.02$ \\
$g = 4$ &  5,000 & $6.82 \pm 0.04$ \\
$g = 5$ &   240 & $4.92 \pm 0.05$ \\
\bottomrule
\end{tabular}
\end{table}

\begin{figure}
\begin{center}
\includegraphics[width=0.4\textwidth]{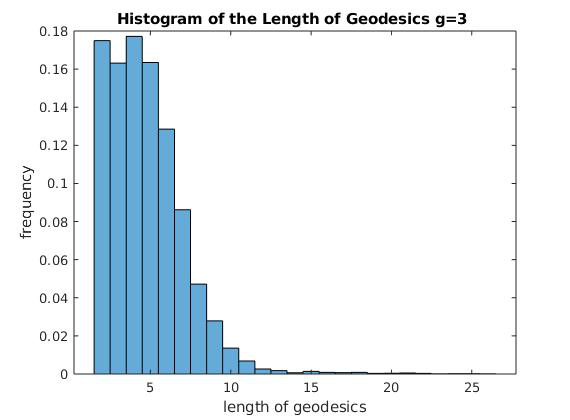}
\includegraphics[width=0.4\textwidth]{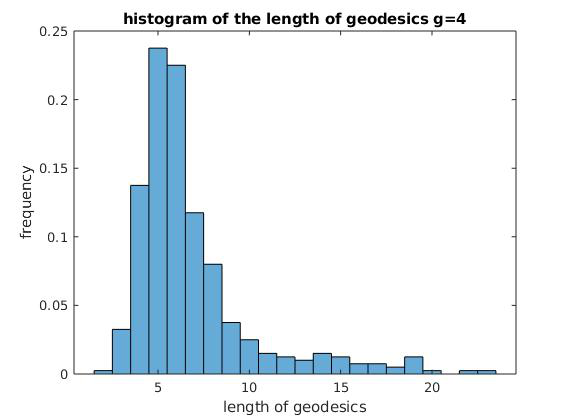}
\caption{Histogram of most likely paths lengths in computations for $g =3$ (left) and $g=4$ (right) genotypes.}
\label{figure3}
\end{center}
\end{figure}

The distribution of lengths for computed geodesics is described by the statistics reported in \tabref{table:geolengths}. We also display the histogram of geodesics lengths for $g=3$ and $g=4$ genotypes in \figref{figure3}. The average length of geodesics for $g=3,4,5$ genotypes is quite short (at the order of $T=6$ days). In the context of \textit{E. coli} laboratory experiments, the doubling time for the colony is approximately 20 minutes under optimal growth conditions. 
However, optimal conditions cannot be maintained for the whole 24-hour period;  the actual growth duration is about 9 to 10 hours per day
(see e.g. \cite{azencoop}),
after that, the population is dormant until the beginning of the next day when a random subsample of fixed size $N$ is generated by dilution, and a fresh daily dose of nutrients is injected.
Thus, in practical experimental setups growth factors are in the range $[200, 240]$.
Therefore, most daily cell lineages involve approximately 70 successive generations of cells and the most likely path of length $T$ corresponds to about $70 \times T$ generations of bacterial cells.

When the terminal histogram $G$ is fairly different from initial histogram $H$, but at the same time has a low concentration of the fittest genotype $k=g$, any most likely path connecting $H$ and $G$ is a rare event occurring with a very small probability. A relatively small most likely path length indicates that whenever a rare event of this type is realized, it essentially happens in a succession of a few highly unlikely steps.  An analogous  phenomenon  has been noted for many diffusions with a small noise (e.g. \cite{ventcell}), where the unlikely escape from a stable equilibrium  point essentially only occurs when the process trajectory picks at each time point the most unlikely direction, namely the direction opposite to the drift.

Next, we consider the $p$-quantile approach for generating the set of penultimate histograms, $\calS$. We consider percentiles 
$p\in\{1\%,5\%,10\%,15\%,20\%,25\%,100\%\}$, and for each pair $(H,G)$ in the random sample we compute the best approximate most likely path using the $p$-quantile algorithm. We then define the efficiency of the $p$-quantile algorithm as the fractions of most likely paths which are identical to the straightforward approach. The efficiency of the  $p$-quantile algorithm is reported in \tabref{efficiencies}.

\begin{table}
\caption{Efficiency of the $p$-quantile algorithm for $g=3,4,5$ genotypes. Efficiency is defined as the fraction of most likely paths coinciding for the $p$-quantile and straightforward algorithms.} 
\label{efficiencies}
\centering
\begin{tabular}{lrrrrrrr}
\toprule
$p$ & 100\% & 75\% & 50\% & 25\% & 10\% & 5\% & 1\% \\
\midrule
$g=3$ & 1 & 1 & 1 & 0.95 & 0.54 & 0.35 & 0.20\\
$g=4$ & 1 & 1 & $1$ & $1$ & $0.94$ & $0.66$ & $0.25$\\
$g=5$ & 1 & 1 & $1$ &$1$ &$1$ &$1$ & $0.71$\\
\bottomrule
\end{tabular}
\end{table}

\begin{table}
\caption{Mean runtime per node in the parallel executions of the most likely path search with the 
$p$-quantile algorithm for the generation of $\calS$.}
\label{runtime}
\centering
\begin{tabular}{lrrrrrrr}\toprule
$p$ & 100\% & 75\% & 50\% & 25\% & 10\% & 5\% & 1\% \\\midrule
$g=3$ & 3 min & 2.5 min & 2 min & 1.5 min & 1 min & 1 min & 1 min \\
$g=4$ & 2 hrs &  1.8 hrs & 1 hr & 7 min & 3 min & 2.5 min & 1.5 min  \\
$g=5$ & 3.2 hrs & 2.2 hrs & 1 hr & 38 min & 32 min & 30 min & 26 min \\
\bottomrule
\end{tabular}
\end{table}

\tabref{efficiencies} demonstrates that the efficiency of the $p$-quantile algorithm increases very quickly as $g$ increases. Our results indicate that the smallest percentile enabling full efficiency verifies $p(3) = 50\%$, $p(4) = 25\%$, $p(5)=5\%$. Thus, for $\zeta=100$, by extrapolating  the observed speed of increase of efficiency as $g$ increases, we conjecture that $p(6) \approx 1\%$, $p(7) < 1\%$, and  $p(g)<<1\%$ for $g\geq 8$. However, the corresponding cardinal number of $\calS$ (same as the number of penultimate histograms to test to compute a single most likely path) can be estimated as $\operatorname{card}(\calS) \approx 5\times 10^4$, $3.2\times10^5$, $5\times10^6$, $10^8$ for $g=3,4,5,6$, respectively. Moreover, $\operatorname{card}(\calS)$ seems to  still increase at an exponential speed  for $g \geq 7$. This is reflected by a significant increase in computational time as $g$ increases reported in \tabref{runtime}. Therefore, this rules out using the $p$-quantiles algorithm for $g \geq 9$, even with efficient parallelization over 20 nodes.

\subsubsection{Simulation Results for $g=7$ and $g=8$ Genotypes}
\label{s:gseveneight}

For a fixed $g$, the efficiency of the $p$-quantile algorithm is an increasing  function of $p$, since the size of penultimate histograms, $\calS$, is an increasing sequence, as $p$ increases. Therefore, we can also define a \textit{relative efficiency}, where we compare the most likely paths computed using two different sets of penultimate histograms, $\calS_{p_1}$ and $\calS_{p_2}$, corresponding to two different percentiles $p_2 > p_1$. The most likely path computed with a higher percentile will always have a lower or equal cost compared to the most likely path computed with a lower percentile because $\calS_{p_1} \subset \calS_{p_2}$.
Since the computation of the most likely paths for $g=7,8$ genotypes is much more expensive compared to $g \le 6$, we analyze the relative efficiency for $g=7,8$ for only one pair of $(H,G)$ and only two percentiles $p_1=10\%$  and $p_2=20\%$.
We consider the mutation rate $m$, the mutation matrix $Q$, and growth factors $F$ as discussed in \apxref{ap:params}. The pairs of histograms $(H,G)$ are presented below.

\begin{description}
\item[Simulations with $g=7$ genotypes:]
\[
H =[0.6, 0.1, 0.1, 0.05, 0.05, 0.05, 0.05], \quad
G = [0.1, 0.1, 0.1, 0.1, 0.1, 0.4, 0.1].
\]
\item[Simulations with $g=8$ genotypes:]
\[
H =[0.5, 0.1, 0.1, 0.1, 0.05, 0.05, 0.05, 0.05], \quad 
G = [0.05, 0.05, 0.1, 0.05, 0.05, 0.1, 0.5, 0.1].
\]
\end{description}

The discretization level for $\calS$ is chosen to be $\zeta=100$ and $\zeta=50$ for $g=7$ and $g=8$, respectively. The $p$-quantile 
algorithm for constructing $\calS$ results in the same most likely path for both $p_1=10\%$ and $p_2=20\%$ and both $g=7,8$. Thus, we conjecture that the efficiency of the $p$-quantile algorithm is close to one for $p=10\%$ and $g=7,8$. The corresponding costs and lengths of the most likely paths for $g=7,8$ are $\operatorname{cost}(\bh^\opt, g=7) = 0.0264$, $\operatorname{cost}(\bh^\opt, g=8) = 0.0534$ and $T^\opt_{g=7} = 6$, $T^\opt_{g=8} = 7$. Most likely paths for $g=7$ and $g=8$ are presented in \tabref{table:geodesicg7} and \tabref{table:geodesicg8}, respectively.

\begin{table}
\caption{The most likely path $\bh^\opt$ connecting $H=H_1$ and $G=H_6$ for $g=7$ genotypes computed with the $p$-quantile algorithm. We display each histogram $H_i$ on the $i$-th row of the table.} 
\label{table:geodesicg7}
\centering
\begin{tabular}{lrrrrrrr}\toprule
\textbf{steps}   & \multicolumn{7}{c}{\bf histogram entries}\\
\midrule
$H_1$ & 0.600 & 0.100 & 0.100 & 0.050 & 0.050 & 0.050 & 0.050 \\
$H_2$ & 0.510 & 0.101 & 0.101 & 0.070 & 0.072 & 0.096 & 0.050 \\
$H_3$ & 0.383 & 0.112 & 0.111 & 0.084 & 0.085 & 0.161 & 0.064 \\
$H_4$ & 0.266 & 0.114 & 0.114 & 0.094 & 0.095 & 0.239 & 0.078 \\
$H_5$ & 0.170 & 0.110 & 0.110 & 0.100 & 0.100 & 0.320 & 0.090 \\
$H_6$ & 0.100 & 0.100 & 0.100 & 0.100 & 0.100 & 0.400 & 0.100 \\
\bottomrule
\end{tabular}
\end{table}

\begin{table}
\caption{The most likely path $\bh^\opt$ connecting $H=H_1$ and $G=H_7$ for $g=8$ genotypes computed with the $p$-quantile algorithm. We display each histogram $H_i$ on the $i$-th row of the table.}
\label{table:geodesicg8}
\centering
\begin{tabular}{lrrrrrrrr}\toprule
\textbf{steps}   & \multicolumn{8}{c}{\bf histogram entries}\\
\midrule
$H_1$ & 0.500 & 0.100 & 0.100 & 0.100 & 0.050 & 0.050 & 0.050 & 0.050\\
$H_2$ & 0.405 & 0.103 & 0.082 & 0.115 & 0.040 & 0.063 & 0.103 & 0.089\\
$H_3$ & 0.301 & 0.101 & 0.095 & 0.112 & 0.046 & 0.077 & 0.171 & 0.097\\
$H_4$ & 0.209 & 0.092 & 0.104 & 0.102 & 0.051 & 0.089 & 0.251 & 0.102\\
$H_5$ & 0.137 & 0.080 & 0.107 & 0.087 & 0.053 & 0.096 & 0.335 & 0.105\\
$H_6$ & 0.085 & 0.065 & 0.105 & 0.069 & 0.052 & 0.100 & 0.420 & 0.104\\
$H_7$ & 0.050 & 0.050 & 0.100 & 0.050 & 0.050 & 0.100 & 0.500 & 0.100\\
\bottomrule
\end{tabular}
\end{table}

For $g=7$ and $g=8$ computational tasks may not be distributed evenly over all parallel nodes due to the requirement to check dynamically whether a particular histogram belongs to the set of penultimate histograms. Therefore, parallelization depends on how the histograms in $\calS$ are distributed over the nodes. Typically, the $p$-quantile is not parallelized well for large $g$, with only a few (between one and five) of the nodes performing the majority of computations. 

We conclude that the $p$-quantile algorithm is feasible and produces adequate results $g\leq 8$, but becomes quite computationally expensive for $g\geq 7$. However, we also conclude that for $g\geq 9$ this strategy is not computationally feasible. Thus, we present next an alternative 
geodesic shooting algorithm based on gradient descent.

\section{Gradient Descent for Reverse Geodesic Shooting}
\label{sec5}

\subsection{Background: Reverse Geodesic Shooting on Riemannian Manifolds}

Let $\mathcal{S}$ be a smooth Riemannian manifold  of dimension $k$. Let us fix the time interval $[0,T]$, as well as two distinct points $x,y\in \mathcal{S}$. As recalled in \secref{analogies}, to seek a Riemannian geodesic $f_t \in \mathcal{S}$ connecting $x$ to $y$, one can solve---under the constraints $f_0=x$ and  $f_T=y$---the Jacobi second order differential equation
\begin{equation}
\left[ \frac{\partial}{\partial f_t} - \frac{\textrm{d}}{\textrm{d}t} \frac{\partial}{\partial f'_t} \right] \mathit{Cost}(f_t, f'_t)  = 0,
\end{equation}

\noindent where
$\mathit{Cost}(f_t, f'_t) = 
\left\langle f'_t, \gamma_{f_t} f'_t \right\rangle$ and $\gamma_z$ is the quadratic form defined by the Riemann metric at $z \in \mathcal{S}$. The Jacobi differential equation satisfied by $f_t$ is then of the form
\begin{equation}\label{e:jacobi}
f''_t = \Gamma(f'_t, f_t) \qquad \text{for} \quad 0<t<T,
\end{equation}

\noindent where $\Gamma(u,v)$ is a fixed smooth function of $u \in \mathcal{S}$, $v \in \bR^k$, easily derived from the Riemann quadratic form $\gamma_z$.

Classical reverse geodesic shooting proceeds as follows: Let $V_y$ be the tangent space to $\mathcal{S}$ at point $y$. Fix any vector $v \in V_y$.  Then solve the ODE~\eqref{e:jacobi} in \emph{reverse time} $t<T$ after initialization by setting $f_T=y$ and $f'_T=v$. Then $f_t = F(t,y,v)$ is a smooth function of $(t,y,v)$. For fixed $x$ and $y$, one  seeks a vector $v\in V_y$  minimizing the Riemannian distance $\mathit{dist}(x,f_0)$ between the prescribed initial  $x$ and $f_0=F(0,y,v)$. This search then proceeds by \emph{numerical gradient descent} in $v$ within the fixed tangent vector space $V_y$.

\subsection{Gradient Descent for Geodesic Shooting in the Space of Histograms}
\label{fixedlength}

Given two interior histograms $H,G \in \calH^+$, we now outline the gradient descent algorithm to compute the most likely path $\bh^\opt$ connecting $H$ and $G$.  The unknown optimal number of steps $T^\opt$ has to be determined as well. We will successively explore the integer values $T=3,4,\dots, T_{\max}$ with the upper bound on the path's length $T_{\max} \leq 5 g$. Let $\calS$ be the set of all histograms $Z \in \calH^+$ such that $\min_{j=1,\ldots,g} Z(j) \geq 0.005$.
Consider a fixed $T \in [3,T_{\max}]$. Then for any $Z\in \calS$ the geodesic shooting algorithm (see \secref{ss:long geo}) computes the \emph{open ended} reverse time geodesic $\{\theta_1= G, \, \theta_2=Z, \dots, \theta_t, \ldots, \theta_{T-1}\}$  by the recursion $\theta_t = \chi(\theta_{t-1}, \theta_{t-2})$, iterated for $t \geq 2$  as long as $\theta_t$ is not a boundary  histogram. 
Then the forward time path $H_t = \theta_{T-t+1}$, $t= 2,\ldots,T \le T_{max}$, is a geodesic connecting histograms $H_2=\theta_{T-1}$ with $H_T=G$ in $T-1$ steps. This path also verifies $H_{T-1}=Z$. We can extend this path to $t=1$ by setting $H_1=H$. This adds a single first step from $H_1=H$ to $H_2\equiv \theta_{T-1}$ with an additional cost $C(H, H_2)$. This extension defines an approximate geodesic $\bh$ connecting $H$ to $G$ in $T$ steps with penultimate histogram $H_{T-1}=Z$. This approximate geodesic is completely determined by $Z$ and $T$.  The total cost of this geodesic is given by
\[
\lambda(\bh)= C(H, H_2)  + \sum_{t=2 \dots T-1} C(H_t,H_{t+1}),
\]

\noindent where $C(y,z)$ is the one-step cost given by~\eqref{e:C(H,G)}. Recall that $C(y,z)$ is a smooth function of interior histograms $y$ and $z$. Moreover one has $H_t = \theta_{T-t+1}$ for $t \geq 2$, where $\theta_t$ is the  smooth function of $Z,G$. Hence, the cost $\lambda(\bh)$ can be treated as a smooth function of the penultimate histogram, $Z$, if we consider $H$, $G$, and $T$ fixed. One can compute the gradient of the total cost  $W(Z) = \partial_Z \lambda(\bh) \in \bR^g$ with explicit formulas summarized in \apxref{s:gradient}. Thus, we can use the gradient descent algorithm to compute the 
optimal path $\bh^\opt$. The gradient descent algorithm results in a sequence of penultimate histograms $Z_1, Z_2, \ldots, Z_n, \ldots$ and  corresponding paths 
$\bh_1, \bh_2, \ldots, \bh_n, \ldots$. We initialize the gradient decent algorithm at random, such that for $Z_1 \in \calS$ the path $\bh_1 = [H, H_{(1),2}, \ldots, H_{(1),T-1}=Z_1, H_{(1),T}=G]$ consists of interior histograms. For $n \geq 2$, gradient descent algorithm  computes 
\[
Y = Z_{n-1} - \alpha_{n} W(Z_{n-1}), \quad
s(Y) = \sum\limits_{j=1}^{g} Y(j), \quad
Z_{n+1}=Y/s(Y),
\]

\noindent where $Z_{n-1}$ is the penultimate histogram from the previous step, computed together with the cost of the path $\lambda(\bh_{n-1})$ and the gradient $W(Z_{n-1})$. The gain parameter $\alpha_n >0$ needs to be chosen small enough to guarantee $\min_j Y(j) > 0$ and also guarantee that the path $\bh_n$ consists of interior histograms. In addition, we use the Armijo line search~\cite{armijo66,Nocedal:2006a,Boyd:2004a} to calibrate the step size $\alpha_n$. The gradient descent algorithm is terminated at step $n$ if either $n=500$ or if the cost $\lambda(\bh_n)$ stopped decreasing. We perform the gradient decent for all values of  $T \in [3,T_{max}]$ and select the most likely path $\bh^\opt$ connecting $H$ and $G$ as the path with the minimal total cost with respect to $T$.

\subsection{Results for Gradient Descent Geodesic Shooting}
\label{sec5.3}

For $g=8$, we tested the gradient descent algorithm for the same pair $H,G$ used in \secref{s:gseveneight} and parameter values in \apxref{ap:params}. For $g=8$, the computational time for the $p$-quantile algorithm discussed in \secref{s:gseveneight} is approximately one hour. The optimal penultimate histogram and the total trajectory cost computed given by the $p$-quantile algorithm are given by the row $H_6$ in \tabref{table:geodesicg8} and $\tilde{\lambda}=0.0534$, respectively.

The gradient descent algorithm yields the most likely path with a smaller overall cost $\lambda^\opt=0.0343$ and the penultimate histogram  
\[
Z^\opt=[0.0841,0.0651,0.1129,0.0664,0.0564,0.0999,0.4188,0.0963].
\]

\noindent  Moreover, the computational time using the gradient descent algorithm  was drastically reduced to 20 seconds on a single computational node. Therefore, the gradient decent algorithm for computing the most likely path is  superior to the $p$-quantile approach discussed earlier in this paper.

Due to a significant reduction in computational time using the gradient descent algorithm,  we are able to apply this approach for $g=10$ genotypes. Parameters are given in \apxref{ap:params}. The initial and the target histograms were chosen as
\[
\begin{split}
& H=[0.45, 0.10, 0.10, 0.05, 0.05, 0.05, 0.05, 0.05, 0.05, 0.05], \\
& G=[0.05, 0.05, 0.05, 0.05, 0.05, 0.05, 0.05, 0.10, 0.45, 0.10].
\end{split}
\]

\noindent The gradient descent algorithm  yields the best geodesic $\bh^\opt = \{H_1 =  H, \ldots, H_9=G\}$ connecting $H$ to $G$ in $T^\opt=9$ steps, with the total trajectory cost $\lambda(\bh^\opt)=0.0259$ and penultimate histogram  
\[
Z^\opt = H_8=[0.0726, 0.0606, 0.0625, 0.0567, 0.0572, 0.0573, 0.0559, 0.0997, 0.3810, 0.0974].
\] 

\noindent The total computing time was around 40 seconds on one single node. Here, we present the analysis of convergence for $T^\opt=9$.

We also studied the rate of convergence of norm $\|H - \theta_{T^\opt-1}\|^2$ and the total cost of the trajectory with respect to the optimization step, $n$. We expect to observe linear convergence since we consider a standard (first-order) gradient descent scheme. We observe that the norm and the cost did not change drastically after approximately $n=18$ for the experiments carried out in this section, showcasing a quick convergence to a good solution for the considered problem (detailed results are reported in \cite{su_thesis}). Since the convergence of gradient descent is in general not independent of the problem dimension, we expect to observe a slower rate at larger values for $g$. Accelerated gradient descent schemes (with momentum) or quasi-Newton methods can serve as a remedy. However, given the overall small dimension ($g \ll 1000$), we do not think higher-order optimization algorithms are required.  

The gradient descent algorithm is considerably faster compared to a more straightforward $p$-quantile discretization approach. In particular, the computing time for $g=10$ only doubles compared with the computing time for $g=8$. Thus, the gradient descent algorithm can be successfully used to analyze the most likely paths for a relatively large number of genotypes. This work will be carried out in a successive paper where we analyze realistic biological scenarios of rare events.

The efficiency and performance of the gradient descent algorithm depend on the initial penultimate histogram, $Z_1$, and selecting an adequate starting penultimate histogram is quite crucial to get the correct most likely path connecting $H$ and $G$. Similar to other applications of the gradient descent algorithm to nonlinear problems, the optimization search can converge to a local minimum. In addition, the gradient descent algorithm is only applicable if the trajectory generated by $Z_1$ consists of interior histograms. Therefore, we performed studies of how random selection of $Z_1$ affects the performance of the gradient descent algorithm. In particular, we generated $K=154$ initial penultimate histograms $\{Z_1^{(k)}, k=1,\ldots,K\}$ and after fixing the number of steps the optimal value ($T^\opt=7$ and $T^\opt=9$ for $g=8$ and $g=10$, respectively), we performed gradient decent search for each penultimate histogram. For $g=8$, 131 paths out of 154 resulted in interior paths. Moreover, approximately 50 paths had a trajectory cost essentially equal to 0.03425, which is approximately the optimal cost for $g=8$. This corresponds to a success rate of  $\approx 38\%$ for random initializations of the gradient descent approach for $g=8$. For $g=10$, only 52 out of 154 random choices of $Z_1$ resulted in paths consisting of interior histograms. Moreover, among these 52 paths, 25 paths had costs in the range
$[0.0216, 0.0235]$ and only 5 had the approximate lowest cost of 0.0216. This implies that it is necessary to perform many realizations of the gradient descent algorithm with many values of the penultimate histogram. It is possible to combine a coarse discretization of the space of interior histograms $\calH$ with the gradient descent algorithm or to find a good starting point by performing a coarse greedy search to find a good initial point.

\section{Conclusions}
\label{sec6}

In this paper, we discuss the large deviations theory for Markov Chains modeling  genetic evolution for bacterial populations. In particular, these models describe the  long-term laboratory \textit{E. coli} experiments where cells undergo daily growth, mutations, and dilutions. It has been demonstrated that such long-term evolutionary experiments often lead to the emergence of new bacterial genotypes. Thus, the Markov chains discussed in this paper are aimed to model the stochastic dynamics of population histograms of genotype frequencies. We outlined the main theoretical results from a previous paper \cite{azencott2018rare, geiger_thesis}. These theoretical results lay the mathematical foundation for deriving  the cost function for evolutionary trajectories connecting two arbitrary histograms. In particular, we focus on rare events and the large deviation analysis of trajectories when one of the emergent bacterial sub-populations does not have the highest fitness (i.e. not the dominant genotype in the population). Such trajectories correspond to the cost-minimizing paths and have very small probabilities which cannot be computed directly by, for instance, ensemble simulations.

We present several algorithms for computing the most likely trajectories  connecting the initial and the final histograms. The computation is performed ``in reverse time'' and the most likely trajectories are completely determined by the final and penultimate histograms. Since the penultimate histogram is not known, this leads to an optimization problem with respect to the penultimate histogram. We present several algorithms for computing the most likely trajectories  connecting the initial and the final histograms. The computation is performed ``in reverse time'' and the most likely trajectories are completely determined by the final and penultimate histograms. Since the penultimate histogram is not known, this leads to an optimization problem with respect to the penultimate histogram. We compare and contrast the straightforward and the $p$-quantiles search algorithms in \secref{s:brute_algo}. These algorithms are based on a relatively straightforward optimization problem, where the set of candidate penultimate histograms is constructed with a relatively coarse discretization. We discuss the efficiency of the $p$-quantile search algorithm and demonstrate its applicability for the number of genotypes $g\le 8$. and the $p$-quantile search algorithms in \secref{s:brute_algo}. These algorithms are based on a relatively straightforward optimization problem, where the set of candidate penultimate histograms is constructed with a relatively coarse discretization. We discuss the efficiency of the $p$-quantile search algorithms and demonstrate its applicability for the number of genotypes $g\le 8$.

To handle problems with a larger number of possible genotypes in bacterial populations, we develop a more efficient gradient descent algorithm in \secref{sec5}. In particular, we derive explicit expressions for the gradient of the cost function with respect to the penultimate histogram. We demonstrate that this algorithm is easily computationally applicable to problems with  $g \leq 10$ genotypes. Moreover, we also estimate that this algorithm is potentially applicable for problems with up to $g=20$ genotypes. Since the corresponding optimization problem is nonconvex, one potential drawback of the gradient descent algorithm is that is  might converge to a local minimum, as discussed at the end of \secref{sec5.3}. Therefore, for problems with $g > 10$ genotypes,  one probably has to combine the gradient descent algorithm with the  $p$-quantiles approach, so that the starting penultimate histogram in the gradient descent algorithm is selected near an optimal point. The computational framework developed in this paper stands ready to tackle interesting biological problems such as the emergence of genotypes with 
lower fitness but with a higher degree of adaptation. This involves considering a mutation matrix $Q$ with non-uniform mutation rates. These experiments will be carried out in a consecutive paper.

Numerical algorithms presented in this paper can be potentially used for two types of applications - (i) reconstructing the most likely path connecting the initial and the final histograms and (ii) performing the importance sampling Monte-Carlo simulations in path space and, thus, computing the 
small probabilities of reaching the target histogram. Both of these applications were addressed in \cite{su_thesis}. In practice, it is not possible to 
determine the genetic composition of the bacterial population during each generation, since these experiments often run for tens of thousands of generations (e.g. \cite{cooper1}). Therefore, the detailed genetic evolutionary path can be reconstructed using the numerical technique presented in this paper. The second application involves performing Monte-Carlo importance sampling simulations where the original Markov chain is modified so that all paths remain close to the most likely path connecting the initial and the final histograms. This enables computing accurate estimates of the actual probability of reaching the target histogram.
We demonstrated that our computational strategy allows addressing both applications in a practical setting with a relatively large number of genotypes $g \approx 20$.

\section*{Acknowledgments} This work was partly supported by the National Science Foundation (NSF) through the grants DMS-2009923, DMS-2012825,  DMS-2145845, and DMS-1903270. Any opinions, findings, conclusions, or recommendations expressed herein are those of the authors and do not necessarily reflect the views of the NSF. This work was completed in part with resources provided by the Research Computing Data Core at the University of Houston.


\appendix

\section{Summary of Notation} Below, we summarize the notation used in this manuscript.
\label{sec:ap1}

\begin{itemize}
\itemsep0em
\item
$g \in \bN$: number of genotypes in the bacterial population
\item
$N \in \bN$: number of cells in the bacterial population
\item
$m \in \bR$: mutation rate
\item
$H \in \bR^g$: population histogram of bacterial frequencies frequencies; $H = [H(1), H(2), \ldots, H(g)]$ with $H(j) \ge 0$
for $j=1,\ldots,g$ and
$\sum_{j=1}^g H(j) = 1$ 
 \item
$H(j) \in [0,1]$: frequency of genotype $j$ in the population
\item
$H_t \in \bR^g$: population histogram of bacterial frequencies on day $t$
\item
$H_t(j) \in [0,1]$: frequency of genotype $j$ in the population on day $t$
\item
$[F_1, F_2, \ldots, F_g]$: ordered genotype growth factors
with $F_1 <F_2 < \ldots <F_g$
\item
$\calH \subset \bR^g$: space of all histograms
\item $\calH^+ =\{H: H \in \calH \text{ and } H(j) > 0 \text{ for all } j=1,\ldots,g\}$ is the 
set of all interior histograms
\item
$\mathbf{H}= \{H_1, \ldots, H_T\}$: time-dependent 
path of length $T$ (days) in the space $\calH$
connecting the initial histogram $H_1$ and the final histogram $H_T$
\item
$\Omega_T$: space of all paths of length $T$
\item
$\Omega_T^+$: space of all interior paths of length $T$; 
if $\mathbf{H} \in \Omega_T^+$ then $H_t(j) \ge \eps$
for all $1 \le t \le T$ and $1 \le j \le g$ and for some
$\eps \ll 1$
\end{itemize}

\section{Growth Factors and Selective Advantages}
\label{ap:params}

In all simulations in this paper the mutation rate is $m=10^{-8}$. The $g \times g$ mutant transfer matrix $Q$ has all diagonal coefficients $Q_{i,i} = 0$, and all non-diagonal coefficients $Q_{i,j} = 1/{g-1}$ for $i\neq j$. 

For simulations with $g =3,4,5$ genotypes growth factors $F = [F_1 , \ldots, F_g]$ 
are given by 
\begin{itemize}
\item 
$g = 3$:  $F = [200, 305.6, 377.7]$,  
\item
$g = 4$: $F = [200, 305.6, 339.7, 377.7]$,
\item
$g = 5$:  $F = [200, 305.6, 339.7, 358.2, 377.7]$.
\end {itemize}

For simulations with $g=7,8$ genotype growth factors are given by 
\begin{itemize}
\item 
$g = 7$:  $F =  [200, 289.8, 305.6, 322.2, 339.7, 358.2, 377.7]$,  
\item
$g = 8$: $F =[200,274.8,289.8,305.6,322.2,339.7,358.2,377.7]$,
\item
$g = 10$:
$F=[200, 200^{1.04}, 200^{1.05}, 200^{1.06}, 200^{1.07}, 200^{1.08}, 200^{1.09} ,200^{1.10}, 200^{1.11}, 200^{1.12}].$
\end {itemize}

\section{Gradients of the Rate Functions}
\label{s:gradient}

\paragraph*{Gradient of Rate Function for Reverse Geodesics.} Any geodesics (represented in reverse time) $\btheta=[\theta_1\equiv G, \theta_2\equiv Z, \ldots, \theta_T=H]$ is fully determined by the final and penultimate histograms $G$ and $Z$, since for $t \geq 3$  $\theta_t$ is given by a recursive relationship $\theta_t=\chi(\theta_{t-1},\theta_{t-2})$  in \eqref{e:reverse}. The trajectory cost is given by 
$\lambda(\btheta) = \sum_{t=1}^{T-1}C(\theta_{t+1},\theta_{t})$ where $C(\cdot,\cdot)$ is the one-step cost 
in \eqref{e:costformula}.  

Next, we consider the number of steps, $T$, fixed and derive the gradient $D_Z\lambda(\btheta)$ with respect to the penultimate histogram $Z$. Using the chain rule
\begin{equation} 
\label{e:Formula5.6}
D_Z\lambda(\btheta)= \sum_{t=1}^{T-1}  \left[
\nabla_x C(x,y) \bigg|_{(x,y) = (\theta_{t+1},\theta_t)} D_Z\theta_{t+1} + 
 \nabla_y C(x,y)\bigg|_{(x,y) = (\theta_{t+1},\theta_t)} D_Z\theta_{t}
 \right],
\end{equation}

\noindent where $\nabla_x C(x,y)$ and $\nabla_y C(x,y)$ are row vectors and $D_Z\theta_{t+1}$ and $D_Z\theta_{t}$ are matrices with $(D_Z\theta_{t})_{ij} = \partial \theta_t(i) / \partial z(j)$. Then $D_Z\lambda(\btheta)$ is a column vector.
We can express the derivative $D_Z\theta_{t}$ as
\begin{equation}   
\label{e:Formula5.7}
D_Z \theta_{t} = 
D_x \chi(x,y) \bigg|_{(x,y) = (\theta_{t-1},\theta_{t-2})} D_Z\theta_{t-1} + 
D_y \chi(x,y) \bigg|_{(x,y) = (\theta_{t-1},\theta_{t-2})} D_Z\theta_{t-2},
\end{equation}

\noindent where both $D_x \chi(x,y)$ and $D_y \chi(x,y)$ are matrices. For $t=1,2$, 
$D_Z \theta_1 \equiv D_Z G = 0$ and 
$D_Z \theta_2 \equiv D_Z Z = I$ (the identity matrix). 
Therefore, formula \eqref{e:Formula5.7} can be used to compute  $D_Z \theta_t$ for $t > 2$. After that, formula \eqref{e:Formula5.6} can be used to compute the 
gradient $D_Z\lambda(\btheta)$. Next, we need to derive expressions for partial derivatives of $C(x,y)$ and $\chi(x,y)$, where $x$ and $y$ are two arbitrary internal histograms.

\paragraph*{Gradient of One-step Cost.} The one-step cost $C(H,G)$ from $H$ to $G$ is given by
\[
C(H,G)=\mathit{KL}(G,\Phi) + m\sum_{j\neq k}F_jH(j)Q_{j,k}(1-U_k/U_j),
\]

\noindent where
\[
\Phi(j)=\frac{F_jH(j)}{\langle F,H\rangle}, \qquad
\mathit{KL}(G,\Phi)=\sum_j G(j)\log(G(j)/\Phi(j)), \qquad
U_j=\exp\left(\frac{G(j)}{F_jH(j)}\right).
\]
Since $H$ is a histogram normalized to one, its entries are \emph{not} independent and we can express $H(g) = 1 - \sum_{j=1 \ldots g-1} H(j)$. Thus, we compute the partial derivatives of $\Phi$ above with respect to $H(j)$ for $j<g$. We obtain the following expressions
\[
\begin{aligned}
\partial_{H(j)}\Phi(j)& =\frac{F_j\langle F,H\rangle - F_jH(j)(F_j-F_g)}{\langle F,H\rangle^2}
&& \text{for}\; j < g,
\\
\partial_{H(k)} \Phi(j) &=\frac{F_jH(j)(F_g-F_k)}{\langle F,H\rangle^2}
&&\text{for}\; j,k < g, j\neq k,
\\
\partial_{H(j)} \Phi(g) & =\frac{-F_g\langle F,H\rangle -F_gH(g)(F_j-F_g)}{\langle F,H\rangle^2}
&&\text{for}\; j < g.
\end{aligned}
\]

Derivatives of KL divergence can be computed as follows
\[
\partial_{H(j)} \mathit{KL} = - \sum_k  \frac{G(k)}{\Phi(k)}\partial_{H(j)} \Phi(k),
\qquad
\partial_{G(j)} \mathit{KL} =\log \frac{G(j)}{\Phi(j)}-\log \frac{G(g)}{\Phi(g)}.
\]

\noindent We can also obtain explicit expressions for the derivatives of $U_j$:
\[
\partial_{H(j)} U_j =- U_jG(j) / F_jH(j)^2  \quad \text{and} \quad \partial_{G(j)} U_j= U_j/F_jH(j) \quad \text{for}\; j < g,
\]
\noindent and
\[
\partial_{H(j)} U_g = U_gG(g) / F_gH(g)^2, \qquad \partial_{G(j)} U_g =- U_g / F_gH(g).
\]

The differential $D_HU$ is a matrix of partial derivatives $(D_HU)_{j,k} = \partial_{H(k)} U_j$. Denote
\[
\rho=\rho(H,G)= m \sum_{j \neq k}F_j H(j) Q_{j,k} U_k/U_j. 
\]

\noindent For $s<g$, the partial derivatives of $\rho$ with respect to $H(s)$ and $G(s)$ are given by
\begin{align*}
\partial_{H(s)}\rho(H,G)
& = m\Bigg[F_g Q_{g,s} \left(-\frac{U_s}{U_g} + H(g)\frac{(D_H U)_{s,s} U_g-U_s (D_H U)_{g,s}}{U_g^2}\right)
\\
& \qquad
+ F_s Q_{s,g}\left(\frac{U_g}{U_s} + H(s)\frac{(D_H U)_{g,s} U_s-U_g (D_H U)_{s,s}}{U_s^2}\right)
\\
&\qquad
+ \sum\limits_{\substack{k=1,\ldots,g-2 \\ k\neq s}} 
\!\!\left(
F_sQ_{s,k}\left(\frac{U_k}{U_s}-\frac{H(s)U_k (D_HU)_{s,s}}{U_s^2} \right) + mF_kH(k)Q_{k,s}\frac{(D_HU)_{s,s}}{U_k}
\right)
\\
& \qquad 
+ \sum\limits_{\substack{k=1,\ldots,g-2 \\ k\neq s}} \!\!
\left(
-F_gQ_{g,k}U_k \left(\frac{1}{U_g}+\frac{H(g) (D_HU)_{g,s}}{U_g^2} \right)+mF_kH(k)Q_{k,g}\frac{(D_HU)_{g,s}}{U_k}
\right)
\Bigg],
\end{align*}

\noindent and
\begin{align*}
\partial_{G(s)}\rho(H,G)
& = m\Bigg[\sum\limits_{\substack{k=1,\ldots,g-1 \\ k\neq s}} \left(- Q_{s,k}\frac{U_k}{U_s}+
F_kH(k)Q_{k,s}\frac{U_s}{U_kF_sH(s)}\right)
\\
& \qquad 
+ \sum_{k=1}^{g-1}\left(Q_{g,k}\frac{U_k}{U_g} - F_kH(k)Q_{k,g}\frac{U_g}{U_kF_gH(g)}\right)\Bigg].
\end{align*}

The formulas above yield the partial derivatives of the cost $C(H,G)$ with respect to $H(j)$ and $G(j)$ for $j < g$,
\[
\begin{aligned}
\partial_{H(j)} C(H,G) & = \partial_{H(j)} \mathit{KL} - \partial_{H(j)} \rho(H,G),
\\
\partial_{G(j)} C(H,G) & =\partial_{G(j)} \mathit{KL} - \partial_{G(j)} \rho(H,G).
\end{aligned}
\]

\noindent These last two formulas then provide the gradients $\nabla_H C(H,G)$ and $\nabla_G C(H,G)$.

\paragraph*{Gradient of Reverse  Geodesic.} As discussed previously, any geodesics $\{\theta_T, \theta_{T-1}, \theta_1\}$ satisfies the recursive relationship given by $\theta_{t}= \chi(\theta_{t-1},\theta_{t-2})$. Thus, we consider partial derivatives of the vector-valued function $x= \chi(y,z)$, where the function $\chi(y,z)$ is defined by the following expressions
\begin{align*}
X_s & =\frac{y_s}{F_s}\exp\left(\frac{F_s}{\langle F,y\rangle}-\frac{z_s}{y_s}\right),\qquad 
Y_s =\frac{X_s}{\sum_t X_t},\\
e_{s,k} & =\exp\left(-\frac{y_s}{F_sY_s}+\frac{y_k}{F_kY_k}\right), \\
\alpha_s & =\sum_{k\neq s}\left(Q_{s,k}e_{s,k}-\frac{F_kX_k}{F_kX_s}Q_{k,s}e_{k,s}\right),\\
f_{s,k} & =\exp\left(-\frac{z_s}{F_sy_s}+\frac{z_k}{F_k y_k}\right).\\
\beta_s & = F_s\sum_kQ_{s,k}-\left(F_s+\frac{z_s}{y_s}\sum_kf_{s,k}Q_{s,k}-\frac{z_s}{F_sy^2_s}\sum_kF_ky_kQ_{k,s}f_{k,s}\right),\\
w_s & =\alpha_s+\beta_s-\langle Y,\alpha+\beta\rangle,\\
x_s & =Y_s + m Y_sw_s.
\end{align*}

For any vector valued function $V= V(y,z) \in \bR^g$ differentials with respect to $y,z \in \bR^g$ are $g \times g$ matrices with entries $(D_y V)_{ij} =  \partial_{y(j)} V(y,z)(i)$ and $(D_z V)_{ij} =  \partial_{z(j)} V(y,z)(i)$. Similarly, the differentials $D_ye$, $D_ze$, $D_yf$, and $D_zf$ of the matrices $e$ and $f$ are $g \times g \times g$ tensors, with coefficients denoted $(D_ye)_{s,k,q} = \partial_{y(q)} e_{s,k}$, with similar notations for  other tensor differentials.

From the expression for $X_s$ we obtain
\begin{align*}
(D_yX)_{s,s}
&=\exp \left(\frac{F_s}{\langle F,y\rangle}-\frac{z_s}{y_s}\right)\left(\frac{1}{F_s}+\frac{z_s}{F_sy_s}-\frac{y_s(F_s-F_g)}{\langle F,y\rangle^2}\right),
\\
(D_zX)_{s,s}
& =-\frac{1}{F_s}\exp\left(\frac{F_s}{\langle F,y\rangle}-\frac{z_s}{y_s}\right),
\\
(D_zX)_{s,k}
& = 0, \quad (D_yX)_{s,k} =\frac{y_s(F_g-F_k)}{\langle F,y\rangle^2}\exp \left(\frac{F_s}{\langle F,y\rangle}-\frac{z_s}{y_s}\right) \quad \text{for } s\neq k.
\end{align*}

Let  $S= \sum_s X_s \in \bR$. Then one has
\begin{align*}
(D_zS)_k &= (D_zX)_{k,k}
\\
(D_yS)_k &=
\exp \left(
\frac{F_k}{\langle F,y\rangle}-\frac{z_k}{y_k}\right) \left(\frac{1}{F_k}+\frac{z_k}{F_ky_k}-\frac{y_k(F_k-F_g)}{\langle F,y\rangle ^2}
\right)
\\
& \quad
+\sum\limits_{\substack{s=1,\ldots,g\\ s\neq k}}
\exp\left(\frac{F_s}{\langle F,y\rangle}-\frac{z_s}{F_s}\right)\frac{y_s(F_g-F_k)}{\langle F,y\rangle^2}.
\end{align*}

Then we can compute
\[
(D_yY)_{s,k} = \frac{S (D_yX)_{s,k} - X_s (D_yS)_k}{S^2}
\quad \text{and} \quad 
(D_zY)_{s,k} =\frac{S (D_zX)_{s,k} - X_s (D_zS)_k}{S^2}.
\]

Next, differentiating $e_{s,k}$ we obtain for $q\neq s \neq k$,
\begin{align*}
(D_ye)_{s,k,s}
& = e_{s,k}
\left(
\frac{y_s (D_yY)_{s,s}-Y_s}{Y_s^2}-\frac{y_k (D_yY)_{k,s}}{F_kY_k^2}
\right),
\\
(D_ye)_{s,k,k}
& = e_{s,k}
\left(
\frac{y_s (D_yY)_{s,k}}{F_sY_s^2}+\frac{Y_k-y_k (D_yY)_{k,k}}{Y_k^2}
\right),
\\
(D_ye)_{s,k,q}
& = e_{s,k}
\left(
\frac{y_s (D_yY)_{s,q}}{F_sY_s^2}-\frac{y_k (D_yY)_{k,q}}{F_kY_k^2}
\right),
\\
(D_z e)_{s,k,q}
& = e_{s,k}
\left(
\frac{y_s}{F_sY_s^2}\cdot (D_zY)_{s,q}-\frac{y_k}{F_kY_k^2}\cdot (D_zY)_{k,q}
\right).
\end{align*}

Differentiating $\alpha$ we obtain for $q<g$
\[
(D_y \alpha)_{s,q}
= \sum_{k\neq q} Q_{s,k} (D_ye)_{s,k,q} - \sum_{k\neq s}\frac{F_k Q_{k,s}}{F_s}\left(e_{k,s}\frac{(D_yX)_{k,q}X_s
- X_k (D_yX)_{s,q}}{X_s^2}+\frac{X_k (D_ye)_{k,s,q}}{X_s}\right)
\]

\noindent and for $s<g$
\[
(D_z\alpha)_{s,s}
= \sum_{k\neq s}\left(Q_{s,k} (D_ze)_{s,k,s}-\frac{F_k X_k Q_{k,s}}{F_s}\frac{(D_ze)_{k,s,s}X_s - e_{k,s} (D_zX)_{s,s}}{X_s^2}\right).
\]

\noindent Next, for $s=1,\ldots,g$, $q<g$ and $s\neq q$,
\begin{align*}
(D_z\alpha)_{s,q}
& = Q_{s,q} (D_ze)_{s,q,q}-\frac{F_qQ_{q,s}}{F_sX_s}
\left(e_{q,s} (D_zX)_{q,q} + X_q (D_ze)_{q,s,q}\right)
\\
&\quad
+ \sum\limits_{k\neq s, \; k\neq q}\left(Q_{s,k} (D_ze)_{s,k,q}-\frac{F_kX_k}{F_sX_s}Q_{k,s} (D_ze)_{k,s,q}\right).
\end{align*}

The expression for $f_{s,k}$ implies that for $s<g$, $k=1,\ldots,g$,
\[
(D_yf)_{s,k,s}= \frac{f_{s,k}z_s}{F_sy_s^2}  
\quad \text{and} \quad
(D_zf)_{s,k,s}=-\frac{f_{s,k}}{F_sy_s},
\]

\noindent and for $s=1,\ldots,g$ , $k<g$, $q \neq s$, $q \neq k$,
\[
(D_yf)_{s,k,k}=-\frac{f_{s,k}z_k}{F_ky_k^2},
\quad 
(D_zf)_{s,k,k} =\frac{f_{s,k}}{F_ky_k},
\quad \text{and}\quad
(D_yf)_{s,k,q} = 0.
\]

The formula for $\beta$ implies that for $s<g$,
\begin{align*}
(D_y\beta)_{s,s}
&= \frac{z_s}{y_s^2}\sum_k f_{s,k}Q_{s,k}-\left(F_s+\frac{z_s}{y_s}\right)\sum_{k\neq s}(D_yf)_{s,k,s} Q_{s,k}\\
&\quad +\frac{2z_s}{F_sy_s^3}\sum_k F_ky_kQ_{k,s}f_{k,s}-\frac{z_s}{F_sy_s^2}\sum_{k\neq s}F_ky_kQ_{k,s} (D_yf)_{k,s,s},
\\
(D_z\beta)_{s,s}
& =-\frac{1}{y_s}\sum_k f_{s,k}Q_{s,k}-\left(F_s+\frac{z_s}{y_s}\right)\sum_k Q_{s,k} (D_zf)_{s,k,s} \\
&\quad - \frac{1}{F_sy_s^2}\sum_k F_ky_kQ_{k,s}f_{k,s}-\frac{z_s}{F_sy_s^2}\sum_k F_ky_kQ_{k,s}(D_zf)_{k,s,s}.
\end{align*}

\noindent For $s=1,\ldots,g$, $q<g$ and $s\neq q$, we have
\begin{align*}
(D_y\beta)_{s,q}
& =-\left(F_s+\frac{z_s}{y_s}\right)Q_{s,q}(D_yf)_{s,q,q}-\frac{z_s F_q y_q Q_{q,s} (D_yf)_{q,s,q}}{F_sy^2_s} -\frac{z_s F_q Q_{q,s} f_{q,s}}{F_sy_s^2},
\\
(D_z\beta)_{s,q}
& =-\left( F_s+\frac{z_s}{y_s})\sum_{k\neq s}Q_{s,k}(D_zf)_{s,k,q} \right)
- \frac{z_s}{F_sy_s^2}\sum_{k\neq s} F_ky_kQ_{k,s}(D_zf)_{k,s,q}.
\end{align*}

The formula for $w$ implies that for $s<g$, $k<g$
\begin{align*}
(D_yw)_{s,k}  & =(D_y\alpha)_{s,k} +(D_y\beta)_{s,k} - (D_yS)_k,
\\
(D_zw)_{s,k}  & =(D_z\alpha)_{s,k} +(D_z\beta)_{s,k}- (D_zS)_k.
\end{align*}

Finally, from $x_s=Y_s+mY_sw_s$, we obtain
\begin{align*}
(D_y x)_{s,k}  & = (D_yY)_{s,k}(1+mw_s)+m Y_s (D_yw)_{s,k},
\\
(D_z x)_{s,k}  & = (D_zY)_{s,k}(1+mw_s)+m Y_s (D_zw)_{s,k}.
\end{align*}

\noindent Since $x=\chi(y,z)$ the last two formulas provide gradients  of $\chi(y,z)$	with respect to $y$ and $z$.


\end{document}